\expandafter\def\csname ver@fixltx2e.sty\endcsname{}
\documentclass[final,5p,times,twocolumn]{elsarticle}

\usepackage{graphicx}
\usepackage{amssymb}

\usepackage[switch]{lineno}

\biboptions{numbers,sort&compress}

\usepackage{svg}
\usepackage{amsmath}
\usepackage{dblfloatfix} 
\usepackage{wrapfig} 
\usepackage{adjustbox}  
\usepackage{multirow}   
\usepackage{url} 
\usepackage{eurosym}
\usepackage{amsfonts} 
\usepackage{dblfloatfix} 
\usepackage{hyperref} 

\usepackage{svg}
\usepackage{amsmath}
\usepackage{dblfloatfix} 
\usepackage{wrapfig} 
\usepackage{nomencl}
\makenomenclature
\usepackage{afterpage}

\usepackage{etoolbox}
\renewcommand\nomgroup[1]{%
  \item[\bfseries
  \ifstrequal{#1}{A}{Abbreviations}
]}
\usepackage[utf8]{inputenc}
\hypersetup{colorlinks=true, linkcolor=blue, anchorcolor=red, citecolor=blue, filecolor=red, urlcolor=red, pdfauthor=author}

\journal{arXiv:2101.10092 [eess.SY]}

\begin{document}

\begin{frontmatter}
\title{Beyond cost reduction: Improving the value of energy storage in electricity systems.}

\newcommand{\orcidauthorA}{0000-0002-4390-0063}
\newcommand{\orcidauthorB}{0000-0002-3494-0469} 

\author[First]{Maximilian Parzen \corref{cor1}\fnref{label2}}
\ead{max.parzen@ed.ac.uk}

\author[Second]{Fabian Neumann}
\author[Third]{Adriaan H. Van Der Weijde}
\author[First]{Daniel Friedrich}
\author[First]{Aristides Kiprakis}

\cortext[cor1]{Corresponding author.}

\address[First]{University of Edinburgh, Institute for Energy Systems, EH9 3DW Edinburgh, United Kingdom}
\address[Second]{Technische Universität Berlin, Department of Digital Transformation in Energy Systems, 10587 Berlin, Germany}
\address[Third]{The Netherlands Organization for Applied Scientific Research (TNO), 2595, DA, The Hague, The Netherlands}

\begin{abstract}

From a macro-energy system perspective, an energy storage is valuable if it contributes to meeting system objectives, including increasing economic value, reliability and sustainability. In most energy systems models, reliability and sustainability are forced by constraints, and if energy demand is exogenous, this leaves cost as the main metric for economic value. Traditional ways to improve storage technologies are to reduce their costs; however, the cheapest energy storage is not always the most valuable in energy systems. Modern techno-economical evaluation methods try to address the cost and value situation but do not judge the competitiveness of multiple technologies simultaneously.
This paper introduces the 'market potential method' as a new complementary valuation method guiding innovation of multiple energy storage. The market potential method derives the value of technologies by examining common deployment signals from energy system model outputs in a structured way. We apply and compare this method to cost evaluation approaches in a renewables-based European power system model, covering diverse energy storage technologies.
We find that characteristics of high-cost hydrogen storage can be more valuable than low-cost hydrogen storage. Additionally, we show that modifying the freedom of storage sizing and component interactions can make the energy system 10\% cheaper and impact the value of technologies.
The results suggest looking beyond the pure cost reduction paradigm and focus on developing technologies with suitable value approaches that can lead to cheaper electricity systems in future.

\end{abstract}

\textbf{Graphical abstract}

\includegraphics[width=1\textwidth]{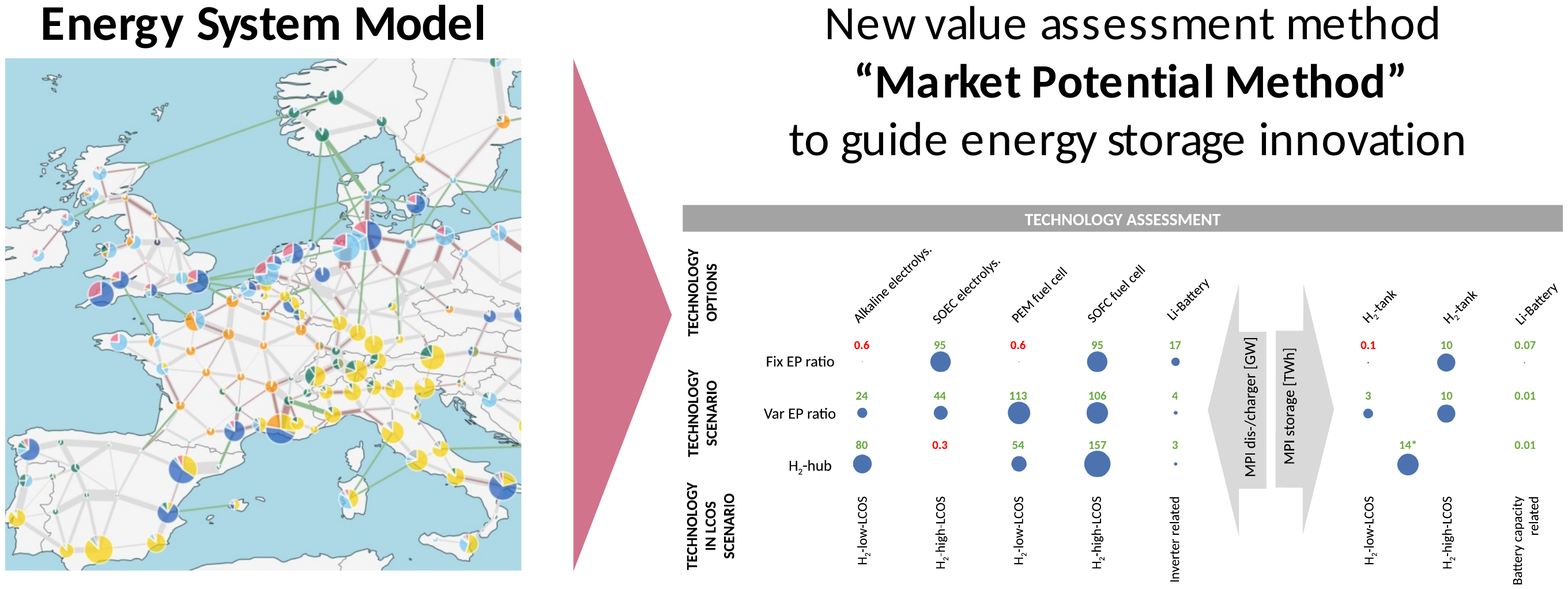}


\textbf{Highlights}
\begin{itemize}
    \item Review of evaluation methods for energy storage identifies need for new approaches. 
    \item Formulation of new 'market-potential method' to identify value of storage.  
    \item Pitfalls of cost approaches are identified in an European electricity system. 
    \item Increasing storage design-freedom impacts technology value and system benefit.  
    \item The 'market-potential-method' is useful for research and industry. 
\end{itemize}


\end{frontmatter}

\section{Introduction}
\label{S:1}

In the face of global ambitions to reduce greenhouse gas emissions, the energy transition characterised by increasing shares of wind and solar power will benefit from more energy storage in the future electricity system \cite{IRENA2017, Umamaheswaran2015, deSisternes2016TheSector}. How many benefits can be delivered by energy storage depends, among others, on how future technology will be designed. Consequently, research and development (R\&D) must evaluate the techno-economic design of energy storage systems to be most beneficial.

A traditional technology evaluation approach is to reduce the cost of its devices \cite{Kittner2017EnergyTransition}. For energy storage, these costs can be defined as absolute costs (\euro), or relative to energy (\euro/kWh) or power (\euro/kW) quantities. In particular, in the material science and chemistry literature, cost reductions of energy storage are a pivotal element, alongside maintaining other storage characteristics such as a 'sufficient' high efficiency, power and energy density, and safety \cite{Lio2015AqueousStorage, Lu2015AStorage}. Though, what is 'sufficient' high is often unclear. Only if one energy storage outperforms the other in all characteristics it represents a superior technology; otherwise, more expensive energy storage with suitable technical characteristics can compete as well (as will be demonstrated in Section \ref{S:Result and Discussion}). Similar, evaluation techniques exist that aim to maximise the profit, however, these are mostly suitable to evaluate single projects (see review in Section \ref{sec:literature-review}). Fortunately, material science literature has recognised one of the key challenges that energy storage depends on different applications and the interaction with the energy system \cite{Liu2013AddressingStorage}. 

Alternative technology evaluation approaches use energy system models. These tools describe energy systems mathematically and capture system-values arising from storage interactions with the wider energy system (see Section \ref{sec:literature-review} for more details). Some studies applying energy system models focus on storage technology evaluation and guidance. For instance, \cite{Sepulveda2021TheSystems} explores the design spaces for long-duration energy storage, \cite{Umamaheswaran2015, deSisternes2016TheSector, Mallapragada2020Long-runGeneration} explore the system-value of generic storage technologies and \cite{Georgiou2020} explores technology specific system-values of liquid-air energy storage and pumped-thermal electricity storage. A limitation of these studies is that counterfactual scenarios constrain this analysis type to single generic or rigid storage examples making the evaluation results questionable.

This study introduces as technology evaluation approach the 'market potential method' which can be described as systematic deployment assessment. Different to classical market potentials that are derived from energy system models which quantify mainly system effects \cite{Cebulla2018HowGermany}, we focus on the systematic assessment of market potentials to evaluate energy storage technologies (see Section \ref{S:market potenial method}). This approach overcomes the previously described limitations and simultaneously analyses multiple and more-flexibly sized energy storage. As we will see later in Section \ref{S:Result and Discussion}, reflecting competitive situations and unique constraint demand and supply mismatches in macro-energy systems are important factors that can affect the system-value of energy storage.

The contribution of this paper to existing literature is as follows:
\begin{itemize}
    \item We review and discuss techno-economic approaches that are currently used to evaluate and compare energy storage technology in Section \ref{sec:literature-review}. We include cost, profit and system-values analysis.
    
    \item We show that current cost metrics can be misleading for technology design decisions. Section \ref{S:LCOS} and \ref{S:MPMapplied} show that a high levelised cost of storage (LCOS) hydrogen storage can be equally or even more valuable than a low LCOS one from the system perspective. We draw this conclusion by observing the deployment of low and high LCOS hydrogen storage systems in a least-cost power system investment planning model. 
    
    \item We extend system-value approaches by the newly developed 'market potential method' in Section \ref{S:market potenial method}. It is further applied and discussed in Section \ref{S:Result and Discussion}. The market potential method systematically evaluates deployment estimations from energy models by looking at a set of probable scenarios in high spatial-temporal resolution over large regions such as Europe. Compared to existing alternatives that are described in Section \ref{sec:literature-review}, the new approach could be potentially more useful and overcomes many limitations. Research and industry could apply the new approach as a complementary tool to guide energy storage innovation.
   
    \item We show that modifying the freedom of storage sizing and component interactions can lead to significant energy system benefits (Section \ref{S:DesignFreedom}) and impact the system-value of a technology (Section \ref{S:MPMapplied}). It underlines the impact of developing and offering adaptive components, such as charger, storage and discharger, separately instead of complete storage systems. 
    
\end{itemize}

In this study, not all energy values are included. In general, energy storage systems can provide value to the energy system by reducing its total system cost; and reducing risk for any investment and operation. This paper discusses total system cost reduction in an idealised model without considering risks. Reducing risk in power systems can be seen as option value \cite{Umamaheswaran2015} leading to a more beneficial investment and operation. Furthermore, only energy balance benefits within a European power system model are included, ignoring other energy sectors apart from the electricity sector. 
This study neglects sub-hourly signals relevant to address grid stability benefits, but includes hourly up to seasonal arbitrage based scarcity signals relevant to address short and long-term balancing benefits (described in Section \ref{subsection:storage scenarios}).

Our findings suggest that a narrow cost focus on designing energy storage is not enough. Future R\&D design decisions should additionally use system-value insights from energy system models. The presented market potential method could be one approach to accomplish this.


\section{Review on Storage Valuation Methods}\label{sec:literature-review}


This section reviews and classifies currently applied storage valuation methods, or in other words, techno-economic analysis approaches that appraise the competitiveness of energy storage including both, technicalities and economic measures. 

\begin{figure*}[t]
\centering
\hspace{-25pt}\includegraphics[trim={-0.5cm 4.5cm 1.5cm 3cm},clip,width=0.9\textwidth]{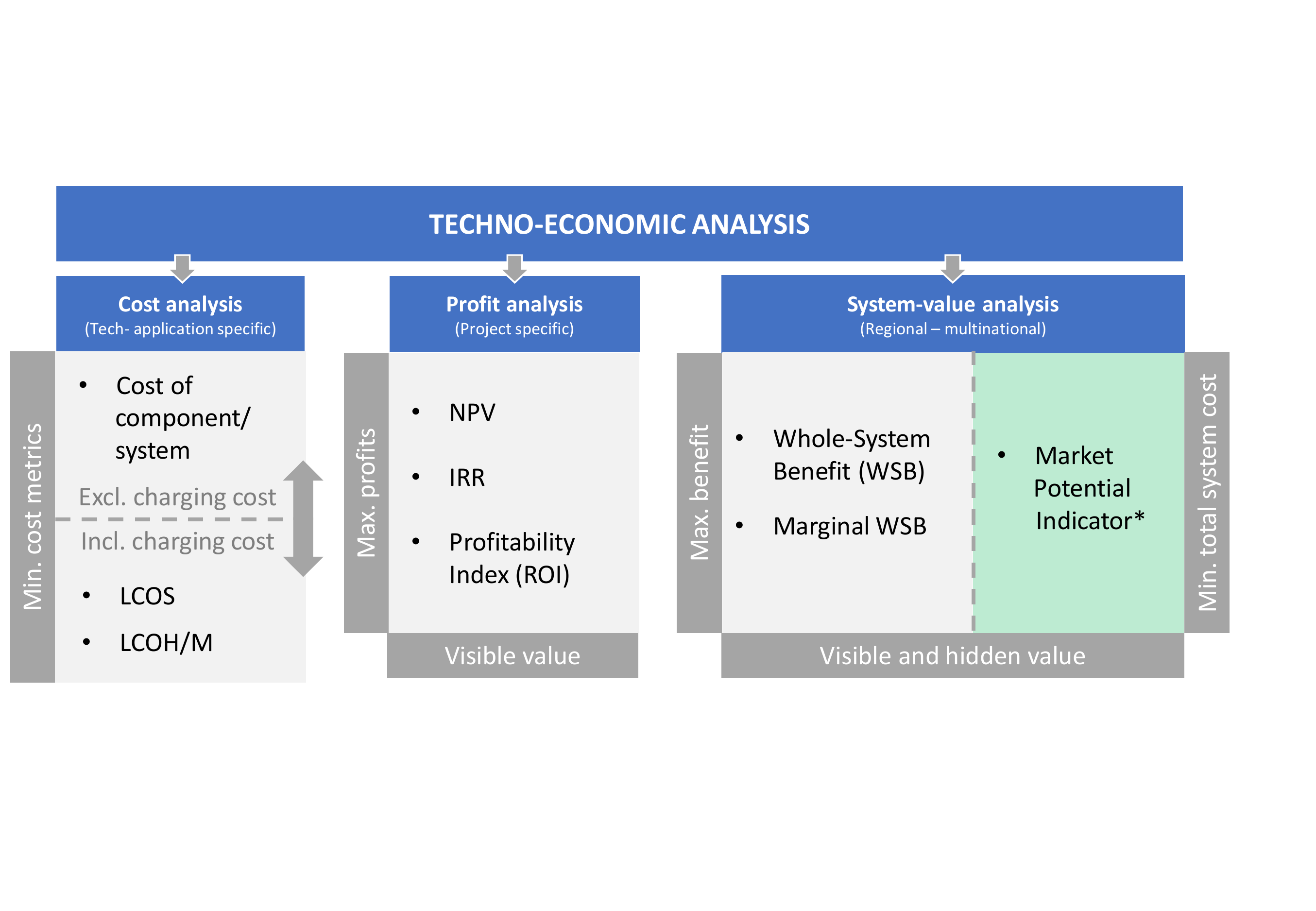}
\caption{Classification of current techno-economic analysis methods in the context of energy storage. *Market potential indicator is a suggested decision metric and part of the newly introduced market potential method. The abbreviations mean the following: levelised cost of storage (LCOS), levelised cost of hydrogen or methane (LCOH/M), net present value (NPV), internal rate of return (IRR), return of investment (ROI). }\label{fig:Classification-Techno-Economic-Analysis}
\end{figure*}

This study classifies the literature into three groups: cost analysis, profit analysis and system-value analysis, which mainly differ in the objective of the metrics. Figure \ref{fig:Classification-Techno-Economic-Analysis} summarises what components will be discussed. These methods are broadly employed for industry decision making, research focus consolidations, and policy regulation \cite{ Schmidt2019, IEA2019, Umamaheswaran2015}, which underlines their importance and the impact of any improvement.  

To understand the 'visible' and 'hidden' value terminology chosen to classify the literature, one should acknowledge that current markets can be considered imperfect and incomplete for multiple reasons: 

\begin{itemize}
    \item Markets are not temporally or spatially resolved. For instance, spot prices are settled over larger spatial areas and not in real-time, leading to not perfect spatial dissolved socialised grid fees \cite{Hirth2015IntegrationVariability}.
    \item Market power can be exploited. Dominant market participants act for their profit while damaging the average participant \cite{Hirth2015IntegrationVariability}.
    \item Forecast information is imperfect. Forecasts of demand, wind and solar generation underlie uncertainties leading to imperfect operation and planning \cite{Hirth2015IntegrationVariability}.
    \item Other negative and positive externalities exist related to incomplete markets, which distort the price. Negative externalities are, for instance, non-priced costs for carbon emission, air pollution and biodiversity losses; positive externalities are non-priced benefits such as non-tracked carbon reduction benefits \cite{Hirth2015IntegrationVariability}. 
\end{itemize}

In this context, system-value analysis generally analyses markets by partially or entirely reducing these market flaws. For instance, energy system models can cover higher spatial and temporal resolution, exclude market power, assume perfect foresight and account for externalities. However, not all models idealise. Some can also incorporate effects of imperfect and incomplete markets by adding cost and benefits related to uncertainty and non-optimal operation and investment \cite{Ringkjb2018ARenewables, Connolly2010ASystemsb, Groissbock2019AreUse}. 

'Visible values' are benefits that can be priced or accounted for in real-world imperfect and incomplete markets as used for profit analysis. In contrast, 'hidden values' are benefits that are not yet priced or accounted for in real world markets. An example are hidden energy storage benefits for network or peak plant deferral or reduced solar and wind power plant curtailments \cite{Sidhu2018AStudy}. To track both hidden and visible values, system-value approaches use idealised models assuming perfect and complete markets. 
 
The following subsections will clarify for each techno-economic analysis class their objectives, methods and users, and further analyse the grade of technical detail and how the approaches handle the role of competition in uncertain future markets.


\subsection{Cost analysis}\label{subsection:Cost analysis}
We categorise the cost analysis of energy storage into two groups based on the methodology used: while one solely estimates the cost of storage components or systems, the other additionally considers the charging cost, such as the levelised cost approaches. Their general objective is to minimise the cost metric for a particular technology or application.

An example of the first approach is represented in \cite{Kassaee2019PARTTechnology}. The energy weighted cost of a storage system (\euro/kWh) is minimised, without any electricity price signal, by a cost optimisation model that simultaneously maximises the round-trip efficiency of the storage. In \cite{DeSantis2017, James2016}, instead of assuming the cost of components, they break down storage components or systems into materials and manufacturing processes. This methodology, known as process-based cost analysis, allows a deeper understanding of cost reductions by mass production or switching to different manufacturing methods. While both approaches do not mention competitiveness or the value of energy storage, their outputs combined with cost and benefit analysis allows finding the value of energy storage solutions.

The levelised cost approaches for energy storage include metrics such as the levelised cost of storage when electricity is discharged (LCOS) and LCOH or LCOM when hydrogen or methane are discharged, respectively \cite{Schmidt2019, Schiebahn2015PowerGermany}. All the levelised cost metrics above are similarly structured. They divide the total cost of the considered system by the discharged energy. Both parameters must be discounted to represent the time value of money \cite{Brealey2020PrinciplesFinance}. Because all levelised cost metrics work similar, we use as generalised form the levelised cost of X (LCOX), where 'X' indicates that the equation holds for various discharged energy carriers:

\begin{equation}\label{eq:LCOX}
\text{LCOX} = \frac{(\sum_0^T \text{Total cost})_{Discounted}}{(\sum_0^T \text{Total discharged energy})_{Discounted}}
\end{equation}

Thereby, the total cost typically consists of capital expenditures, operational expenditures and charging expenditures \cite{IRENA2019, Julch2016ComparisonMethod, Lazard2018}. Sometimes additional factors are included that can impact total cost and total discharged energy, such as degradation rates, taxes, or self-discharging \cite{Schmidt2019}. 

Levelised cost metrics are used to evaluate many applications, such as energy arbitrage, frequency regulation, voltage regulation, system restoration and operational management (i.e. redispatch). For this purpose, the levelised cost metric assumptions must be categorised for the specific application, such as charging price, operational time and power to energy ratio \cite{Schmidt2019, Lazard2018}. 

While the 'cost of component' or 'cost of system' approach is widely used for design decisions with high technological detail \cite{Kassaee2019PARTTechnology,DeSantis2017, James2016}, the levelised approaches forego some technical detail to inform project developers and policy about their projected competitiveness in the market \cite{Schmidt2019}.

Cost of component or system metrics are excellent for exploring cost reduction opportunities in great technical detail. On the other hand, LCOS-like metrics differ by being a good first indicator for the competitiveness between various technologies for a particular application. 

A technology improvement should lead to total system cost reductions. However, the main limitation of cost-analysis methods is that cost reductions for one energy technology can be only a clear signal for technology improvement under the condition that its other techno-economic characteristics do not degrade. For example, an energy store only clearly improves if the cost reduces at least for one component such as charger, store or discharger, while the other component costs and efficiencies are not negatively influenced. If this is not the case, a complex solution space exists for which a more costly energy storage can lead to lower total system cost, and hence, being more valuable, see Section \ref{S:Result and Discussion}. 


\subsection{Profit analysis}

The profit analysis describes methods from the investor's perspective. They tend to choose profitable energy storage projects at current energy market designs \cite{Albertus2020Long-DurationTechnologies, Jiang2019SizeDecisions}. Thereby, the general objective for the investor is to maximise the profit indicator for a given investment. 

The inclusion of discharging behaviour and revenue streams are distinctive for profit analysis. Depending on the market design, several different revenue streams for energy storage exist. In the UK, for instance, 14 potential revenue streams exist, such as frequency response provision or wholesale market arbitrage, which can be power (\euro/kW) or energy (\euro/kWh) related \cite{Jones2016CrackingThem.}. In general, not every storage has access to the same revenue streams due to specific characteristics and requirements \cite{Schmidt2019}. Most studies include only the energy arbitrage service from energy storage, which means buying cheap electricity and selling it later more expensive \cite{Anderson2017Co-optimizingMarkets}. Other studies co-optimise multiple energy services, which result in higher benefits \cite{Anderson2017Co-optimizingMarkets, Das2015AssessingGrid, Kirli2020Techno-economicActions}.

The profit analysis typically evaluates energy storage projects with capital budgeting techniques based on discounted cash flow methods to acknowledge the time value of money \cite{Brealey2020PrinciplesFinance}. The energy storage literature uses multiple project assessment metrics: present value (PV) is employed to calculate the feasible cost of a storage project \cite{Albertus2020Long-DurationTechnologies}, net present value (NPV) to evaluate the profitability of a project \cite{Sidhu2018AStudy, Glenk2019EconomicsHydrogen}, and internal rate of return (IRR) to determine at which discount rate or opportunity cost a project is viable \cite{Anderson2017Co-optimizingMarkets, Walker2016EconomicMechanisms}. NPV and IRR are good investor signals when investment capital can be accessed easily. However, when investment capital is limited, projects should be evaluated by a profitability index, which relates the discounted benefits to the cost \cite{Brealey2020PrinciplesFinance}. Many energy storage studies, therefore, investigate energy storage by the profitability index \cite{Brealey2020PrinciplesFinance}, which is also termed cost-benefit ratio \cite{Denholm2017EvaluatingPlants, Dufo-Lopez2015Techno-economicStorage}, NPV-ratio \cite{Bartela2016Technical-economicInstallation}, return of investment (ROI) \cite{Naumann2015Lithium-ionApplication}, return on equity (ROE) \cite{Jiang2019SizeDecisions}, all giving the signal of how much money can be achieved per investment. Another common metric in the context of energy storage is the payback period \cite{Khosravi2018EnergySystem, Walker2016EconomicMechanisms,Atherton2017Techno-economicFarms}, which \cite{Brealey2020PrinciplesFinance} judges to be an illustrative but not useful factor for investment decisions. Finally, when multiple energy storage technologies with different lifetimes are evaluated and compared, such as in \cite{Dufo-Lopez2015Techno-economicStorage, Glenk2019EconomicsHydrogen, Atherton2017Techno-economicFarms}, an equivalent annual annuity metric is recommended \cite{Brealey2020PrinciplesFinance}. For instance, one could break down the NPV to an equivalent annual annuity where the highest annuity is the preferable project.

The main limitation of the profit analysis is that it misses the 'hidden' or broader power system cost and benefits of energy storage. Because it only focuses on the 'visible' cost and benefits at the current market design. Future energy markets might internalise 'hidden' benefits, such as shown in market design efforts to address the previously hidden greenhouse gas emission costs. Hidden costs and benefits are, for instance, savings due to investment deferral of network upgrades or peak plants, or when fewer curtailments increase the value of renewable generators \cite{Hirth2013ThePrice}. Employing a hybrid method of profit and system-value analysis, the authors in \cite{Sidhu2018AStudy} added social or 'hidden' benefits to the NPV metrics, which are not directly accounted for in the market design. This lead to a higher value of energy storage solutions. The drawback of the approach is that many assumptions are made and added exogenously to the NPV characteristics ignoring the spatial and temporal heterogeneity of the hidden cost and benefits. What may be a reasonable assumption at one location at a specific time must not be the case at another location at the same or another time. Including these variables endogenously, as some energy system models do, can help anticipate better infrastructural changes and reduce risks.

As a result, the profit analysis is a useful method to investigate a storage project's value and competitiveness at present for a specific location at current market designs. This might be sufficient for investors to assess short-term projects at specific locations. However, when one looks at the value of energy storage in the long term or across many regions, the following system-value approach can give some extra insights. 

 
\subsection{System-value analysis}

As previously stated, the system-value analysis estimates the value of energy storage which are 'visible' and 'hidden' at existing markets, for longer time horizon and large spatial regions by considering perfect and complete markets in the analysis. Energy system models are used for the system view, which optimises investment and operation of generators, networks and storage or demand response units at the same time to accomplish the objective of minimising total system cost. The results of such analysis are nowadays mainly applied for policy recommendations. However, they also reveal insights for technology design. For instance, it was found that high capacity factor wind turbines can be equally desired in an optimal energy system as their less capital intensive alternative technology with lower capacity factors -- having smaller hub heights and shorter blade lengths \cite{Hirth2016System-friendlyPower, DallaRiva2017ImpactsEurope}. 

The system-value approaches are important to identify the benefits of energy storage. Which benefits are considered depends on the energy system model design. For instance, \cite{deSisternes2016TheSector} neglects network expansion, missing significant network expansion cost savings from storage deployment \cite{Umamaheswaran2015}. On the contrary, the authors in \cite{Georgiou2020, Umamaheswaran2015} use a model that incorporates generation, network, and system operations savings from energy storage in the UK. 

The whole-system benefit (WSB) given in \euro/year and the marginal WSB given in \euro/kW or \euro/kWh are two inspiring concepts how to attach a system-value to the energy storage in power systems \cite{Umamaheswaran2015, deSisternes2016TheSector, Mallapragada2020Long-runGeneration, Sepulveda2021TheSystems}. Both concepts share a comparison of a none or existing storage scenario with one that includes an energy storage expansion. Such approaches are also known as counterfactual scenarios \cite{Bistline2020EnergyNeeds}. Thereby, the total system cost difference between the scenarios is the WSB that the energy storage creates \cite{Georgiou2020}. When the marginal WSB curve, given in \euro/kW or \euro/kWh, is integrated by the respective storage unit (in kW or kWh), then the WSB is obtained. 
The marginal WSB is described as vital since it provides the upper-cost limit for energy storage for a given amount of installed storage \cite{Strbac2012StrategicFor}. Only if the marginal value is above its marginal cost, the storage is an economically viable option and should be installed. Additionally, to the WSB and its marginal value, the authors in \cite{Strbac2012StrategicFor} extended the concept by the differentiation of the benefits in net and gross benefit. The gross benefit excludes the investment cost of energy storage, while the net benefit includes them. Thereby, the gross value method is used to benchmark how much the cost can rise for a given technology. The net benefit analyses the holistic value for a specific storage case. 

Both WSB methods above lead to insightful results. For instance, (i) that every additional installed energy storage capacity decreases its marginal value; (ii) that the value of energy storage can suffer from competition with other flexibility providers, such as demand response or bi-directional charging of electric vehicle; and finally (iii) that energy storage benefits can be decomposed into its origins such as network and peak capacity savings \cite{Umamaheswaran2015, Georgiou2020}.

The drawback of the WSB approaches is that they are unsuitable as evaluation metrics to signal between multiple storage alternatives what technology is more competitive. The WSB approaches seem to work correctly only for a single energy storage design. When multiple energy storage units are included in the WSB analysis at the same scenario and with variable sizing for each location, it becomes difficult with counterfactual approaches to allocate benefits. Or, in other words, it becomes unclear which energy storage at what location is responsible for certain energy storage benefits at a specific time. As a result, WSB approaches cannot assign a value to one particular storage or compare multiple storage technology candidates.

In the next section, the 'market potential method' aims to extend the existing system-value literature to circumvent the above issue and give decision-maker signals even under complex competition situations. In short, the new approach moves away from assigning monetary values directly to individual energy storage units but instead focuses on the optimised quantity, which means that a storage is likely to be valuable when a certain amount of storage is built. As in Section \ref{S:MPMusefulness} discussed, the quantity appears to be another helpful metric for industry and research when systematically applied. 


\section{Methodology}\label{S:Methodology}
The methodology section is built up as follows. First, the new system value assessment method, the 'market potential method' is defined in theory. Second, an experimental model setup for hydrogen and battery storage is described that compares cost and system-value analysis approaches. Finally, to carry out the experiment, the power system model PyPSA-Eur is introduced with its problem formulation, set of scenarios and model input data.

\subsection{Market potential method} \label{S:market potenial method}

\begin{figure}[h]
\centering
\hspace{-25pt}
\includegraphics[trim={0cm 9cm 10cm 1.5cm},clip,width=0.5\textwidth]{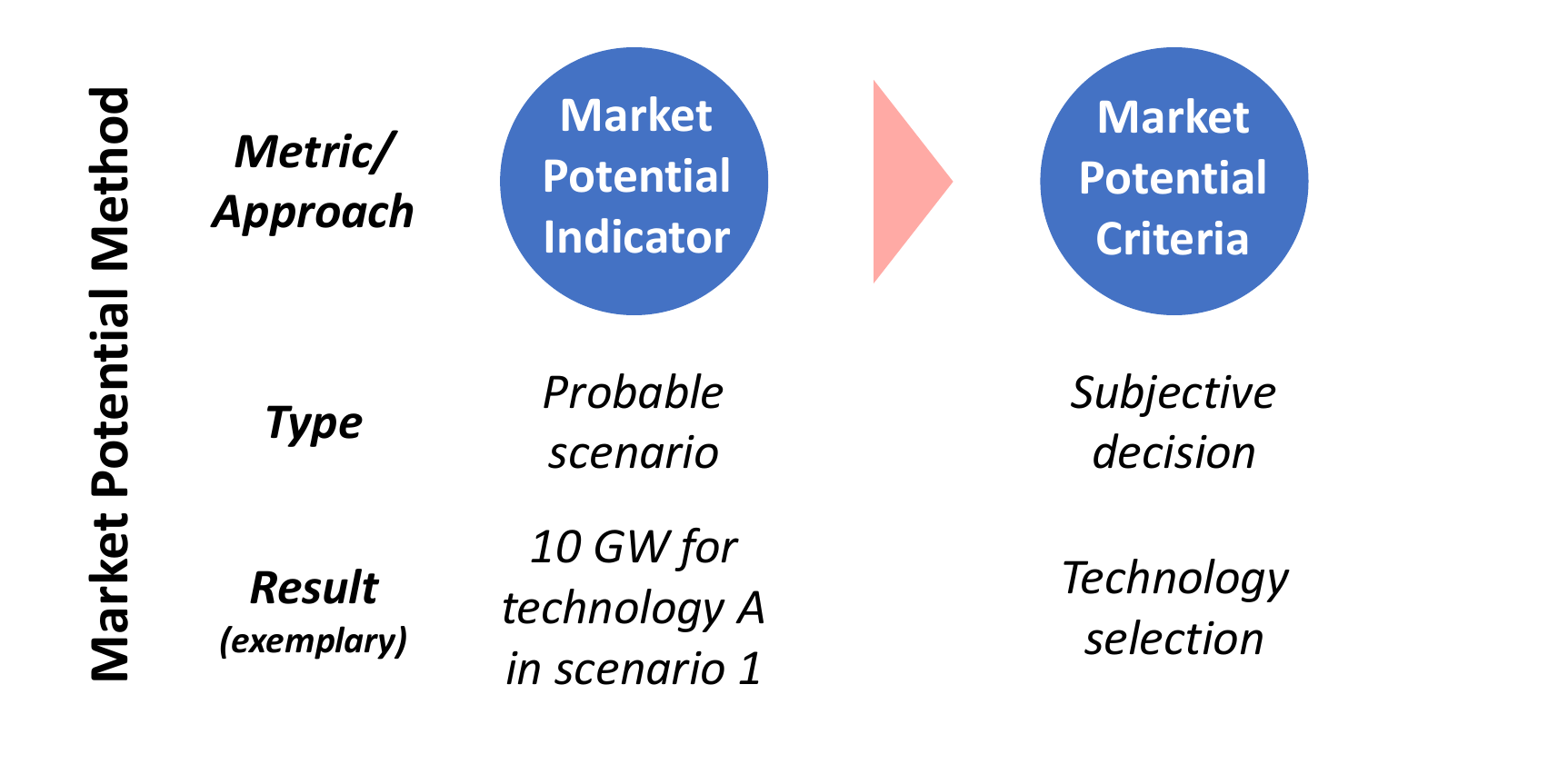}
\caption{High-level description of the Market Potential Method. First a market potential indicator is derived for a single or multiple possible scenarios. The market potential indicator is then used by an entity through a market potential criteria to support design-decisions making on energy storage technology.  }
\label{fig:MarketPotentialMethod}
\end{figure}

The 'market potential method' attempts to expand the existing system-value methods to give more useful signals of which storage technology is valuable in existing or future energy systems. Figure \ref{fig:MarketPotentialMethod} illustrates that the 'market potential method' consists of: first, the 'market potential indicator', which corresponds to the expanded power or energy capacities of a storage component such as charger, discharger or capacity unit; second, the 'market potential criteria' which seek to support design-decision making of storage technologies. 

\subsubsection{Market potential indicator}
The foundation of the introduced method is the market potential indicator (MPI).
The MPI is not a new metric. It is a result of energy system models that analyse scenarios in future energy systems and describes the total quantity of a particular storage technology in a cost minimised electricity system  \cite{deSisternes2016TheSector, Schill2018Long-runSensitivities, ENTSO-E2020ENTSO-E2020}. However, the MPI has never been a central metric to improve, compare and explore storage designs in detail; it was rather used to inform policymakers and market participants about probable energy futures to reduce investors risk \cite{ENTSO-E2020ENTSO-E2020}. We utilise the MPI to guide technology innovation with probable scenarios and market potential criteria.

The market potential can be either aggregated or disaggregated. In the context of energy system models, we define the disaggregated MPI of a storage unit as optimised (or expanded $t-t_0$) power or energy-related size at a region. Thereby, the market potential focuses on the storage component $c$, representing a charger, discharger or store unit. The over a region $i$ aggregated MPI is determined by:

\begin{equation}\label{eq:market potential}
\text{MPI}_{t-t_0,c} = \sum_{i\in\mathbb{N}} \text{(MPI)}_{t-t_0,c,i}   \hspace{7mm} [MW\hspace{1mm}or\hspace{1mm}MWh]
\end{equation}

It is crucial to consider the MPI by components rather than by a fixed-sized storage system for mainly two reasons. First, grid-scale energy storage can be highly scalable and adaptable \cite{Dowling2020RoleSystems, Beuse2020ProjectingSectorb}. For instance, electrolysers (MW), steel tanks (MWh) and fuel cells (MW) composing hydrogen storage systems can be freely scaled and combined. Moreover, in a $H_2$-hub operation, two different electrolysers could feed the same $H_2$-storage tank. Second, energy storage system components--for instance, hydrogen--are not required to be at one location. Indicated by \cite{Schiebahn2015PowerGermany}, hydrogen pipelines can become an economically viable option when large amounts of hydrogen need to be transported. Its integration means that hydrogen electrolyser and fuel cell are not required to be located in one place. Consequently, because storage components can be independently scaled, adaptable in operation and do not require co-location, it seems advisable to optimise them separately.

\subsubsection{Scenario selection and dealing with uncertainty}\label{S: scenario selection}
The use of energy system models is subject to uncertainty as predicting the future with certainty is impossible. It is impossible because we can make decisions that impact the future, such as done by agreeing on multilateral $CO_2$ targets, which improved renewable energy deployment and led to learning by doing cost reductions effects \cite{Kittner2017EnergyTransition}. Nevertheless, analysing a broad range of future scenarios can reduce uncertainty \cite{Schnaars1987HowScenarios}. 

The market potential method in linear programming models relies on possible and probable scenarios. Many different ways exist to create 'possible' scenarios which differ in the set of deterministic input assumption and constraints \cite{Schnaars1987HowScenarios,Nielsen2007Energy303}. However, a possible future does not necessarily mean that it is a probable one. A good approach to develop scenarios that can be expected in future is to follow the ones which are provided and encouraged by either national or multinational institutions - and engage in public consultations if they require changes \cite{ENTSO-E2020ENTSO-E2020}. An example of the latter one is the European Network of Transmission System Operator for Electricity (ENTSO-E) which provides updates on multiple pathway scenarios every two years based on storylines towards the European agreed targets - known as Ten-Year Network Development Plan (TYNDP) \cite{ENTSO-E2020ENTSO-E2020}. Transparency in energy modelling, also from trusted institutions, is a key requirement to lower uncertainty \cite{Pfenninger2018OpeningLearned}.

Scenarios can be additionally selected to investigate multiple technology designs. For instance, technology manufacturers might be interested in such analysis to guide energy storage innovation. 

This study includes three different hydrogen design constraints and two different charger and discharger technologies for technology assessment, which are described in more detail in Section \ref{subsection:storage scenarios}. While this study uses an exemplary 100\% GHG emission reduction scenario that is sufficient for the research purpose, future work should include probable scenarios such given by national or multinational institutions like ENTSO-E.

\subsubsection{Market potential criteria}\label{MPC}
The 'market potential criteria' give the market potential indicator its meaning and can help with decision-making. The criterion includes two simple rules. In an optimised energy system model with many if not all technological alternatives, the technology with:

\begin{itemize}
    \item $MPI = 0$, for one scenario is probably not valuable.
    \item $MPI > 0$, for one scenario is probably valuable.
\end{itemize}

Additionally, the positive MPI magnitude can be used as supportive decision criteria to deal with uncertainty. This can be, for instance, the 'threshold' or the 'bigger is better' rule described below:  

\begin{itemize}
    \item $MPI > X$ or 'threshold rule'. Where a company or institution decides what minimum market potential $X$ must be achieved. For instance, an alkaline electrolyser needs to have a market size of 1 GW to be an attractive technology for a company.
    \item $MPI_A > MPI_B$ or 'bigger is better' rule. If two technologies A and B are compared, the one with higher market potential is more likely to be valuable.
\end{itemize}

In particular, when the evaluation condition appears in multiple scenarios, it reduces the uncertainty of the statements. For instance, when hydrogen storage is significantly optimized in all scenarios it is a clear indicator that it is likely that the technology is valuable in many different probable futures.

\begin{figure}[h]
\centering
\includegraphics[trim={1cm 15cm 10cm 0.5cm},clip,width=0.48\textwidth]{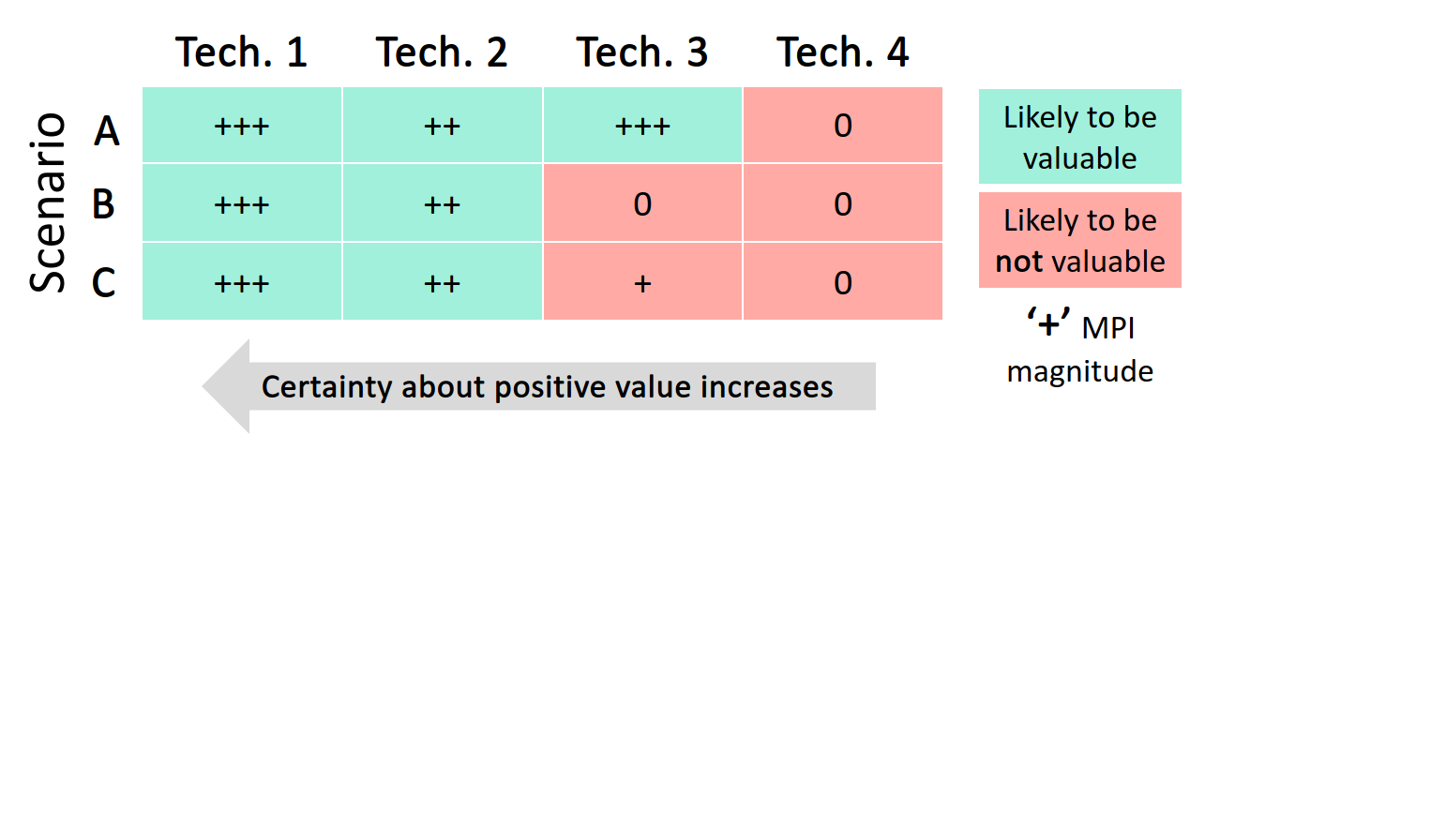}
\caption{Qualitative illustration of market potential criteria applied to a set of scenarios and technology options. The "+" indicates the MPI magnitude. Additionally, the threshold rule is set to a single plus, meaning that a company requires at least two plus to consider a technology as a potential candidate to manufacture or start R\&D activities.}
\label{fig:market-potential}
\end{figure}

Figure \ref{fig:market-potential} illustrates how the market potential criteria could be applied as a decision support tool. The illustrative example could lead to the anticipative decision of a technology manufacturer or research institution to focus rather on the first two technologies than the latter ones.

Only with the criteria one can systematically analyse the market potential indicators and reduce risk. Together, the market potential indicator and criteria build the market potential method. 

\subsection{PyPSA-Eur. Model structure and data} \label{Model structure and data}

The open European transmission system model PyPSA-Eur is adopted to determine the value of various energy storage systems in a European electricity system. PyPSA-Eur is an adaptable investment and dispatch model built on the core model PyPSA that combines high spatial and temporal resolution. The suitability of PyPSA-Eur for operational studies and long-term power system planning studies is described in \cite{Horsch2018PyPSA-Eur:System, Brown2018PyPSA:Analysis, Groissbock2019AreUse}. This section briefly introduces the model structure and applied data. The full model formulation of PyPSA-Eur is given in the Appendix.

\begin{figure*}[t]
\centering
\includegraphics[trim={0cm 5cm 1cm 4cm},clip,width=0.8\textwidth]{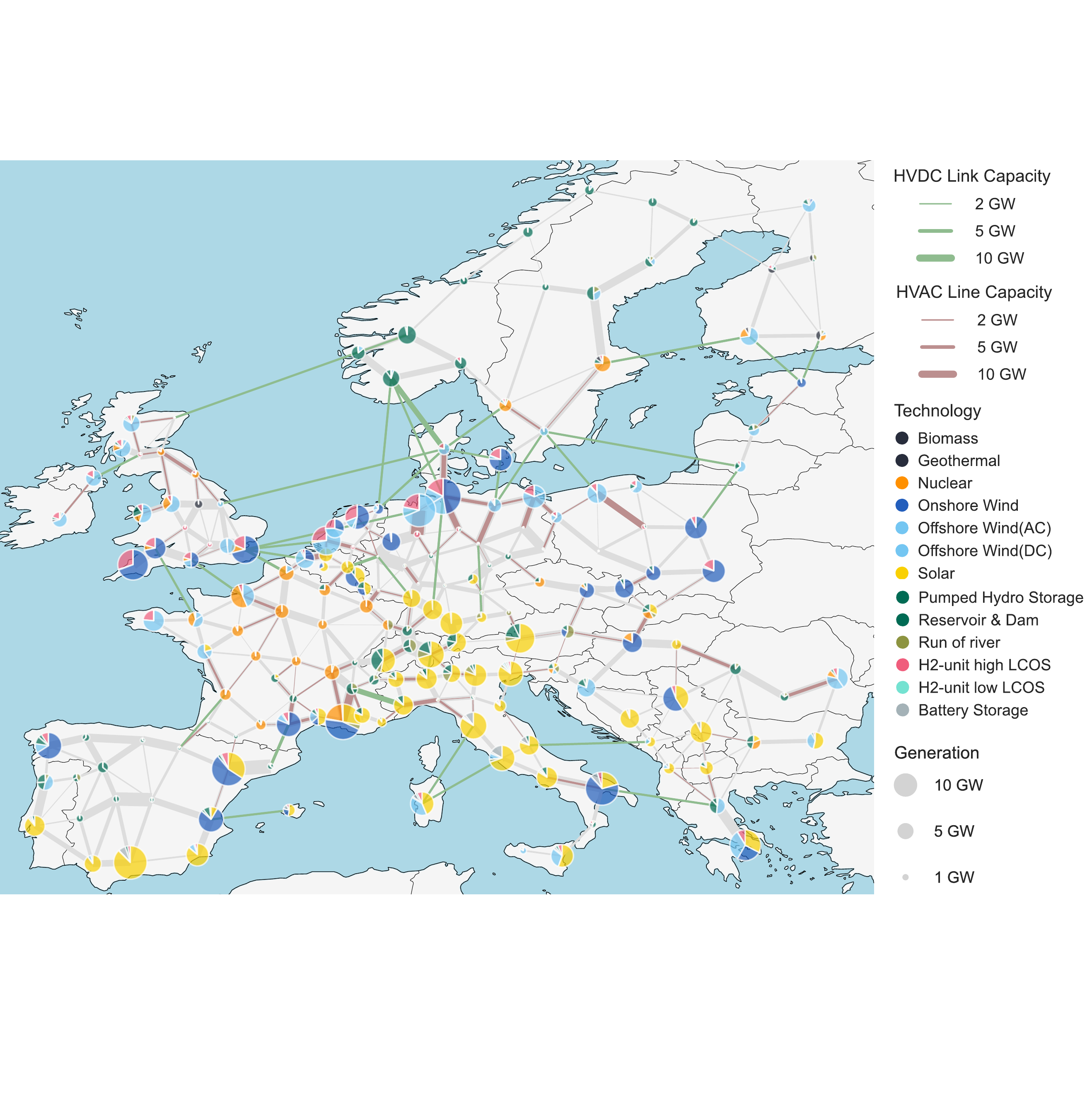}
\caption{Optimal generation, storage and network expansion under a 100\% emission reduction scenario and technology data for 2030. Light grey lines showing the existing installed network capacity, dark grey lines the additional expanded capacity. Plot produced with PyPSA-Eur.}
\label{fig:EU-map}
\end{figure*}

PyPSA-Eur covers the European transmission model and processes electricity system data from diverse sources. Existing conventional generators, transmission lines, substations, and hydro storage systems, as well as planned network reinforcements, are included with their size and location. Wind and solar based technologies are greenfield optimised, which means that existing solar and wind capacities are disregarded. The time series for wind and solar generators are derived from satellite and earth observatory data \cite{Horsch2018PyPSA-Eur:System}. Regarding power demand, the load time series are collected from ENTSO-E data for each country and redistributed by GDP and population over the regions. A spatial resolution of 181 nodes matched with an hourly resolution across an entire year accounts for the complex spatio-temporal patterns of renewables and grid congestion events that shape investment decisions \cite{Frysztacki2021TheSolar}. 

In terms of market economics, the model assumes perfect competition and foresight for one reference year. A detailed model description and formulation is included in \cite{Horsch2018PyPSA-Eur:System, Neumann2021CostsSystem, Neumann2021TheModel, Brown2018PyPSA:Analysis}. Here, we only highlight the key features and constraints. The model's objective is to minimise the total system cost in the European electricity system at the transmission level. The total system costs consist of

\begin{itemize}
    \item investment costs, which includes the annualised capital cost of onshore and offshore wind turbines, storage components and both HVAC and HVDC transmission lines, and
    \item operating costs, which includes fixed operation and maintenance, and variable operating cost. 
\end{itemize}

The objective is subject to

\begin{itemize}
    \item nodal power balance constraints that guarantee that supply equals demand at all times,
    \item linearised power flow constraints modelling the physicality of power transmission, 
    \item Solar and wind resource constraints that limit the theoretical generation time-series. We chose a single weather year for our analysis; however, this can be extended for a more robust prediction of weather year anomalies or variations \cite{Staffell2018TheDemand}.
    \item Renewable availability constraints which restrict solar and wind technical potential based on environmental protection areas, land use coverage and distance criteria.
    \item Emission constraint introduces a limit of carbon dioxide $CO_2$ equivalent emission in the model that impacts technology investment and generation.
\end{itemize}

The model has many adjustable constraints. This study, similar to many others such as \cite{Poncelet2020UnitFlexibility}, does not include the available unit commitment (UC) constraints. In fact, UC constraints are becoming increasingly negligible in future energy systems with increasing shares of renewables and energy storage. Mainly, because it was observed that they only have minor impacts on investment and operational outcomes \cite{Poncelet2020UnitFlexibility}. Further, UC constraints introduce extra computational burdens by the mixed-integer formulation, which removes model convexity and, hence, leads to a nonlinear program that requires more efforts for solving. Therefore, we decided to exclude UC constraints due to their minor impact on the results and large impact on the already heavy computational requirements for the optimization (8 cores, 180 GB RAM solved for roughly 13h with Gurobi). Nevertheless, if a more detailed technological performance in a high renewable electricity system with flexibility constrained nuclear power plants is essential, this UC formulation could be included.

For the input cost and technical assumptions, the documented dataset provided in \cite{Horsch2020PyPSA-Eur:Code} is used, referring to an electricity system scenario in 2030. We only adjusted the dataset of \cite{Horsch2020PyPSA-Eur:Code} by the battery and hydrogen storage system inputs summarised in Table \ref{Tech-details-storage-1} and Table \ref{Tech-details-storage-2}.
\begin{table}[h]
\centering

\caption{Power related energy storage model inputs representing 2030 data
}\label{Tech-details-storage-1}
\centering

\begin{adjustbox}{width=\linewidth}
\begin{tabular}{ l c c c c c c c c}
 
 \hline 
Energy storage components & \multicolumn{2}{c}{Electrolysor}  & \multicolumn{2}{c}{Fuel cell}  & \multicolumn{2}{c}{Battery Inverter}\\
 \hline
LCOS Scenario & [Low]&[High]& [Low]&[High]& [-]\\
 \hline
 \hline
Investment \([EUR/kW_{el}]\) &339&677&339&423$^b$&209$^c$\\ 
FOM$^a$ \([\%/year]\) &2&3&2&3&3\\
Lifetime \([a]\) &25&15&20&20&10\\
Efficiency \([\%]\) &68&79&47&58&90\\
Discount Rate \([\%]\) &7&7&7&7&7\\
 \hline
\multirow{2}{*}{Based on Ref.} &\cite{IEA2019}&\cite{IEA2019}&\cite{Steward2009ScenarioMedium}& \cite{Steward2009ScenarioMedium, Brown2018SynergiesSystem} & \cite{Brown2018SynergiesSystem, Mongird2019EnergyReport}\\ 
&Alkaline & SOEC$^d$ & PEM$^e$ & SOFC$^f$ & Li-Ion Battery$^g$ \\ 

\hline 
\hline
 \multicolumn{7}{l}{$^a$ Fixed operation and maintenance cost}  \\
 \multicolumn{7}{l}{$^b$ Includes fuel cell stack replacement after 10 years which cost 30\% of initial cost} \\
 \multicolumn{7}{l}{$^c$ Includes 80 $EUR/kW$ balance of plant, mainly assigned to wiring and connection \cite{Mongird2019EnergyReport}}\\
 \multicolumn{7}{l}{$^d$ Solid-Oxide Electrolyser}\\
 \multicolumn{7}{l}{$^e$ Proton Exchange Membrane or Polymer Electrolyte Membrane}\\
 \multicolumn{7}{l}{$^f$ Solid-Oxide Fuel Cell}\\
 \multicolumn{7}{l}{$^g$ Lithium-Ion Battery}
  
\end{tabular}
\end{adjustbox}
\end{table}

\begin{table}[h]
\centering
\caption{Energy related energy storage model inputs representing 2030 data}\label{Tech-details-storage-2}
\centering

\begin{adjustbox}{width=\linewidth}
\begin{tabular}{ l c c c c c c c c c c}
 
 \hline 
Energy storage components & \multicolumn{2}{c}{\(H_2\) storage} & Battery storage\\
 \hline
LCOS Scenario & [High]&[Low]& [-]\\
 \hline
 \hline
Investment \([EUR/kWh_{el}]\) &8.4&8.4&188$^b$\\ 
FOM$^a$ \([\%/year]\) &-&-&-\\
Lifetime \([a]\) &20&20&10\\
Efficiency \([\%]\) &-&-&-\\
\hline
\multirow{2}{*}{Based on Ref.} &\cite{Brown2018SynergiesSystem}&\cite{Brown2018SynergiesSystem}&\cite{Mongird2019EnergyReport}\\ 
& \multicolumn{2}{c}{$H_2$ steel tanks} & Li-Ion Battery \\
\hline 
\hline
\multicolumn{4}{l}{$^a$ Fixed operation and maintenance cost}  \\
\multicolumn{4}{l}{$^b$ Includes 81 $EUR/kW$ for engineering, procurement and construction costs \cite{Mongird2019EnergyReport}} 
\end{tabular}
\end{adjustbox}
\end{table}


\subsection{Energy storage scenarios}\label{subsection:storage scenarios}

This study looks at three different constraint energy storage scenarios in one fully emission-free energy system scenario. As explained in Section \ref{S: scenario selection}, one energy system scenario is just exemplary chosen and sufficient for this research. Multiple system scenarios from trusted organisations such as ENTSO-E should be applied if technology decisions are made with the MPM. As mentioned in \cite{Hirth2016System-friendlyPower}, the energy technology impacts the system value, however, the energy system layout and constraints also impact he technology value. Therefore Section \ref{S: scenario selection} goes through the main scenario design elements, the energy system and storage scenario design. 

Starting with the energy system layout and constraints, Figure \ref{fig:EU-map} shows an example of the optimised European electricity landscape for the variable energy-to-power ratio scenario, which is minimised in terms of total system costs in a 181 bus spatial resolution. One should note that the network structure is based on ENTSO-E data which is aggregated to show realistic line capacities between the buses.

Different to \cite{Victoria2020EarlyOff}, the scenarios include the existing European nuclear power fleet but acknowledge the German, Spanish, Belgium and Swiss nuclear exit. The inclusion of nuclear power plants reduces the required VRE capacity expansion and, at the same time, increases the share of dispatchable power plants -- a measure that reduces energy storage demand. However, the flexibility of nuclear plants is overestimated in this study as typical ramp rates reaching up to 36\%/h and minimum allowable power of 20\% per nominal power \cite{Cany2018NuclearExperience} are ignored. However, we ignore such unit commitment constraints to keep the model formulation convex and reduce the amount of variables for computational speed (see more details in Section \ref{Model structure and data}). It implies that this study will tend to underestimate the energy storage potential.

Further, similar to \cite{Neumann2021CostsSystem}, an equity constraint is included that requires every country to produce at least 80\% of its total electricity demand, leading to a smooth distribution of generators across all of Europe. This constraint is motivated by the fact that political leaders avoid depending entirely on electricity imports but are willing to trade considerable amounts to handle the trade-off between the economic benefits of importing cheaper electricity and the sometime costly independence of supply such for isolated networks.

\begin{figure}[h]
\centering
\includegraphics[trim={0.5cm 5.9cm 7.5cm 0.1cm},clip,width=0.48\textwidth]{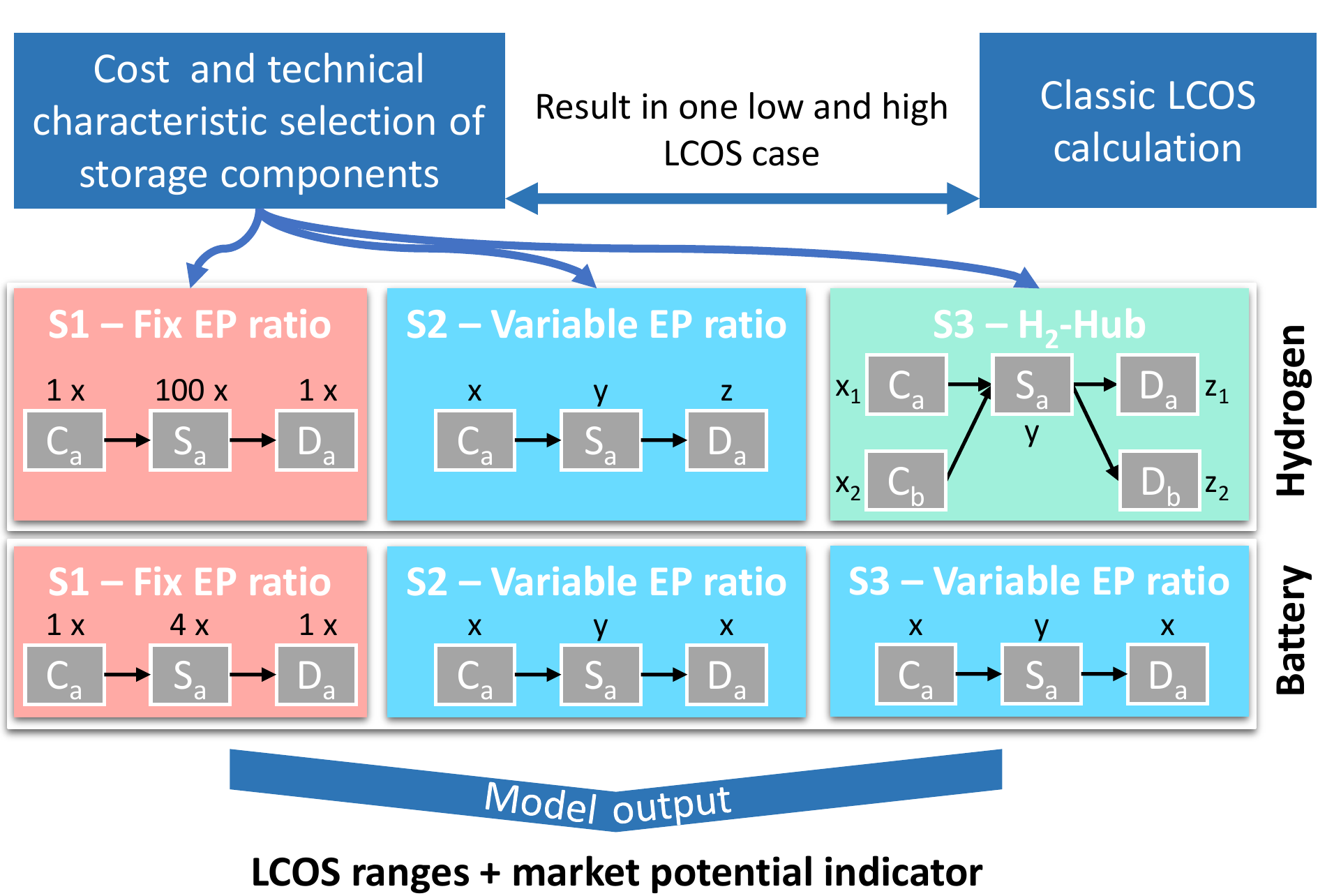}
\caption{Description of the three storage scenarios. The cost and technical storage parameters are chosen once and serve as input for all storage scenarios. Scenario 1 shows the fixed energy-to-power ratio of the hydrogen and battery unit $a$. In Scenario 2 and 3 all components can be freely scaled. However, the battery is constrained to the same charger to discharger ratio. Further, the '$b$' in the $H_2-Hub$ scenario indicates a new technology addition. A least-cost optimization is run with each scenarios, whose results are used to create the spatially resolved LCOS and market potential signals.}\label{fig:scenario-set-up}
\end{figure}

The network expansion is constrained to a volume of $25\%$ compared to the existing network capacity, acknowledging the increasing political difficulty to develop new transmission lines. A limited network expansion can potentially lead to higher storage demand \cite{Neumann2021TheModel}. Further constrained are hydro storage technologies. While these are based on actual power plant data, no further capacity expansion is allowed due to natural limitations in most regions.

The energy storage scenario design is described in Figure \ref{fig:scenario-set-up}. First, technical and economic parameters are chosen as model input for each storage component (see Table \ref{Tech-details-storage-1} and Table \ref{Tech-details-storage-2}) to represent a low and high levelised cost of storage (LCOS) case for classical LCOS calculations. Afterwards, the resulting techno-economic details are inserted in the model environment into three scenarios. The scenarios differ mainly in technological design freedoms. 'Fix EP ratio' is the most constrained energy storage scenario having a fixed energy-to-power ratio of 100 h for the hydrogen and 4h for the battery storage technology -- such as applied in a similar range in research \cite{Denholm2019TheStates, Schmidt2019, Albertus2020Long-DurationTechnologies}. Similar to previously mentioned research publications, this fix EP scenario also assumes that charger and discharger size are equally sized. Otherwise, 'Variable EP ratio' optimises for the hydrogen storage unit each component size, charger, storage and discharger so that the energy-to-power ratio is variable. Here, the battery remains constrained in flexible sizing as charger and discharger represent the same component, namely the inverter, so that the battery storage can only size inverter and battery capacity related design separately (see Battery component size variables $x,y,x$ in Figure \ref{fig:scenario-set-up}). While both fix and variable EP ratio scenario optimise low-LCOS and high-LCOS hydrogen components separately, the '$H_2$-Hub' scenario permits cross operation of hydrogen technologies. This can be considered a $H_2$-Hub, having at one location techno-economically different low and high LCOS charging and discharging technologies that operate the same hydrogen storage. After applying the scenarios in the optimization, the model results are used to create the spatially resolved LCOS and market potential signals which are further discussed in Section \ref{S:Result and Discussion}.

This study creates energy storage scenarios that focus on energy arbitrage benefits under spatially resolved perfect and complete markets. Scarcity signals relevant to seasonal balancing are considered through 'unconstrained' locational marginal prices, also known as nodal prices. These nodal prices can increase to extremely high prices such as more than 20000\euro/kWh and let energy storage be optimised as a seasonal reserve, shifting cheap energy of one season to times of high prices. As introduced in Section \ref{sec:literature-review}, the complete market considerations include the often unaccounted or 'hidden' values of energy storage systems, such as:

\begin{itemize}
    \item Avoided investment cost of network expansion
    \item Avoided investment and operational cost of dispatchable generators
    \item Increased power plant utilisation/ less curtailment 
\end{itemize}  

Emission targets play for the energy storage market potential a vital role. To keep the comparability between scenarios and a decent amount of market potential for energy storage, we set in all scenarios the $CO_2$ emission reduction target to 100 \%.


\section{Results and Discussion}\label{S:Result and Discussion}


\subsection{Relaxing design constraints of energy storage and its benefits}\label{S:DesignFreedom}

As introduction to the cost and value analysis scenarios, this section discusses the impact of design freedom on the storage components and the total system.

Increasing design freedom of energy storage can lead to significant benefits in the electricity system. When investigating the competitiveness of energy storage, many studies assume that the energy to power ratio is fixed \cite{deSisternes2016TheSector, Julch2016ComparisonMethod}. However, assuming a fix energy to power ratio on a continental scale is an unrealistic extreme as well as assuming that all market participants choose the perfect sizing for the market. 

\begin{figure}[ht]
\centering
\includegraphics[trim={0cm 0.2cm 0cm 0cm},clip,width=0.45\textwidth]{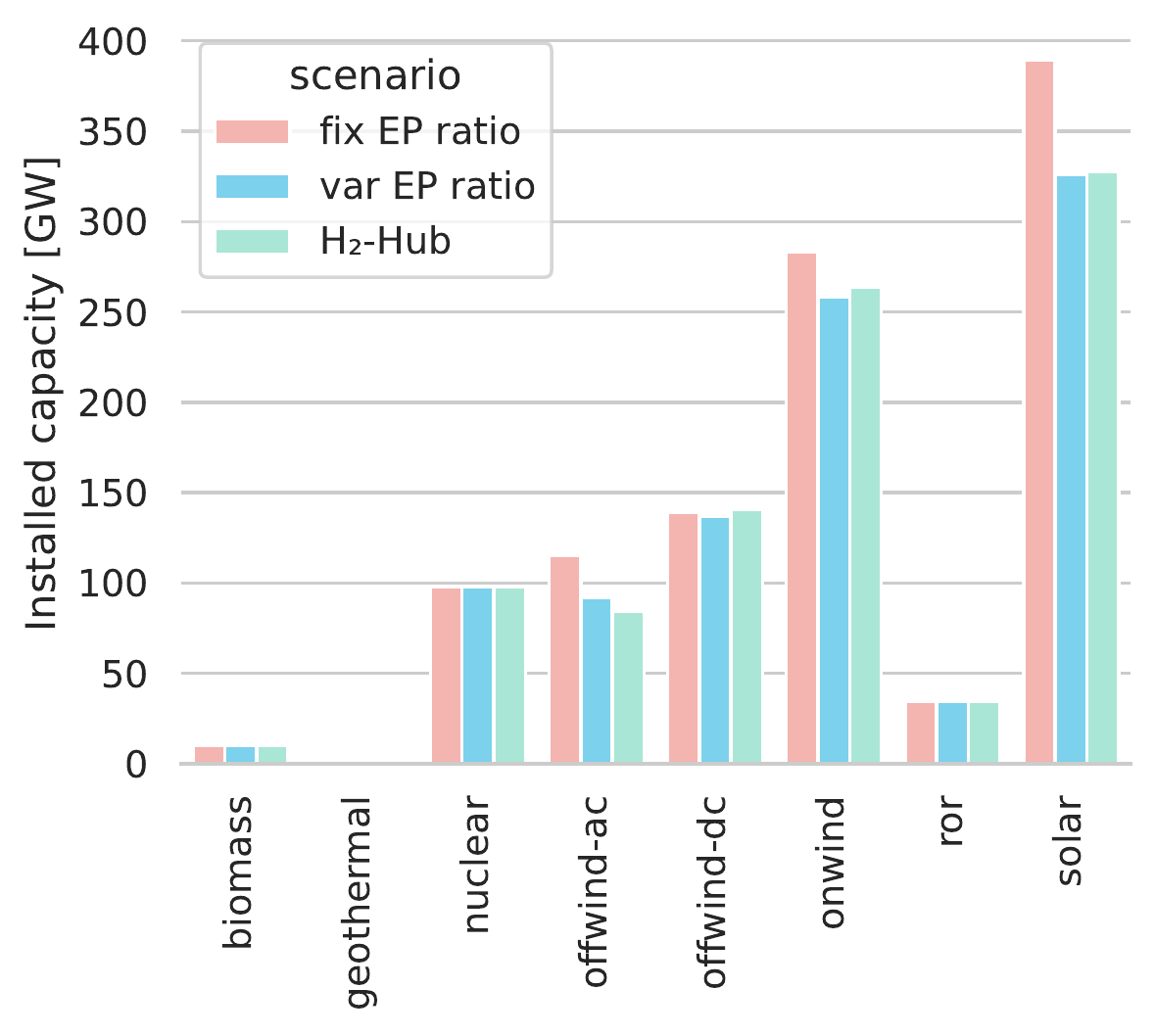}
\caption{Optimization result for future installed generation capacity in the exemplary 100\% emission reduction scenarios. The abbreviations 'ror' stands for run of river, offwind-ac and -dc for AC and DC connected offshore wind plants, respectively.}
\label{fig:installed generator capacity}
\end{figure}

Table \ref{Total system costs} shows that the increasing sizing complexity, however, seems worthwhile to consider as it can lead to per annum total system cost savings of approximately $13B$\euro or $10\%$ in the modelled zero $CO_2$ electricity system scenario while not leading to significant generation portfolio changes (see Figure \ref{fig:installed generator capacity}). Looking at the generation portfolio, the optimization result are representing currently installed power plants in the EU for nuclear, biomass and run-of-river \cite{Gotzens2019PerformingDatabases}. We prohibit these technologies from additional expansion to replicate political constraints. That is why they are not increasing in volume. Similar, these technologies are not decreasing in volume because they are optimized and, hence, desirable options in the given least-cost scenarios. While geothermal is allowed for expansion it does not expand in future scenarios. This indicated that the technology does not contribute to the least cost optimization result for the existing cost assumptions in the power only scenario. Note that this result might change when changing assumptions or adding sectors such as heating and cooling.

The total system cost thereby includes the optimisation relevant costs, which consist of newly installed generation, storage and network components, including any operational costs. Another approach to comprehensively quantify the savings is by calculating the relative investment cost, which divides the total system costs by the total electricity demand. It shows that the introduction of optimised sizing can lead to electricity bill savings of roughly half a cent, with the $H_2$-Hub scenario contributing only to negligible more savings. As a result, increasing design freedom of energy storage can be desirable for a cheaper electricity system and should be considered while designing technology.

\begin{table}[htp]
\centering

\caption{Annual total system costs, relative investment and curtailment data. Variable sizing of energy storage reduces the system costs by 10\%.
}\label{Total system costs}
\centering

\begin{adjustbox}{width=\linewidth}
\begin{tabular}{ l c c c}
 
 \hline 
\multirow{2}{*}{Scenario} & \multirow{2}{*}{Total system cost} & \multirow{2}{*}{Relative investment$^a$} & Curtailment \\ &&&[\% of annual demand]\\
 \hline
 \hline
Fix EP ratio    & 152.9 B\euro &   4.874 ct/kWh & 0.61\% \\
Var EP ratio    & 139.9 B\euro &   4.460 ct/kWh & 0.73\% \\
H2-hub          & 139.7 B\euro &   4.453 ct/kWh & 0.37\% \\

\hline 
\multicolumn{4}{l}{$^a$ Total system cost per annual demand}  \\
\end{tabular}
\end{adjustbox}
\end{table}


The optimal storage design depends on location and technology. Figure \ref{fig:EP_ratio} shows the EP-ratio for multiple locations and technologies with relevant market potential in an optimal European future scenario. 

Hydrogen chargers are smaller sized, and reveal a wider span of EP-ratios than their discharger opponents, which means that slow charging and quick release seem to be beneficial from an EU system perspective at most locations. Further, the Li-Ion batteries are optimised with a 2-4 h EP-ratio, much smaller than the hydrogen components. The reason for that heterogeneous design is that local diverse electricity system situations with its network constraints, supply and demand curves, as well as the different storage characteristics (see Table \ref{Tech-details-storage-1} and \ref{Tech-details-storage-2}) benefit from a variety of storage scaling to reach an optimal solution that minimises the electricity bills.


\subsection{Static LCOS vs modelled LCOS}\label{S:LCOS}

The LCOS is currently an influential metrics to benchmark technology and to discuss their competitiveness. Therefore it is not surprising to see that technology design is even optimised for minimum levelised costs (see Section \ref{sec:literature-review}). To show the drawbacks of this measure, static and modelled values are calculated according to the methodology described in Equation \ref{eq:LCOX}.

The main difference between static and modelled LCOS is what assumptions are used. The static LCOS calculation uses directly assumed or exogenous variables such as for full load hours, electricity prices and energy-to-power ratios. In contrast, the modelled LCOS is based on endogenous variables determined by the energy system model and its inherent assumptions. It means that full load hours, electricity prices and energy-to-power ratios are determined for each location by the European power system model.

The static LCOS is calculated with the technical and economic component characteristics in Table \ref{Tech-details-storage-1} and \ref{Tech-details-storage-2}, and the LCOS assumptions given in Table \ref{table: additional model inputs for LCOS}. The results of the static LCOS calculation also given in Table \ref{table: additional model inputs for LCOS} show a 19.2\% or 5 ct/kWh difference for the two hydrogen storage units, whereby the battery storage seems much more competitive.

In contrast, the modelled LCOS results are given in Figure \ref{fig:LCOS-Ranges} for most buses in the EU electricity system for the 'variable EP ratio' scenario. Despite having the same input cost, lifetime, discount factor and efficiency data as the static LCOS calculation, a wide LCOS range can be observed for each optimised storage unit which consists of charger, storage, discharger. The LCOS ranges are roughly between 20-100, 20-55 and 4-14 ct/kWh for the low, high LCOS $H_2$ unit and the battery. One reason for the wide LCOS ranges is the heterogeneous charging and discharging behaviour, which is indicated by diverse full load hours observed between 80-3000h; another one, the heterogeneous nodal prices or electricity price profiles at each region; and, finally, the heterogeneous sizing of the storage chain. While the battery technology seems more competitive under the LCOS framing, it becomes ambiguous for hydrogen with the overlapping LCOS ranges.

\begin{table}[t]
\centering
\caption{Additional inputs for LCOS calculation oriented on \cite{Schmidt2019} and \cite{Albertus2020Long-DurationTechnologies}}\label{table: additional model inputs for LCOS}
\centering
\begin{adjustbox}{width=0.48\textwidth}
\begin{tabular}{ l c c c}

\hline
 &\multicolumn{2}{l}{Hydrogen storage} & Battery storage\\
 \hline
LCOS scenario &[Low]& [High] &[-]\\
\hline
\hline
Discharging ratio \([h]\)&100&100&4\\
Electricity price \([Eur/MWh]\)&50&50&50\\
Yearly full load hours \([h]\)&2500&2500&3400\\
Roundtrip efficiency$^a$ \([\%]\)&32.0&45.8&81,0\\
Lifetime \([a]\)&25&15&10\\
\hline
Static LCOS$^b$ \([ct/kWh]\)&0.21&0.26&0.12\\
\hline
\hline
\multicolumn{4}{l}{$^a$ calculated product from energy storage component efficiencies in Table \ref{Tech-details-storage-1}}\\
\multicolumn{4}{l}{$^b$ calculated with Equation \ref{eq:LCOX}, and inputs from Table \ref{Tech-details-storage-1} and \ref{Tech-details-storage-2}, \ref{table: additional model inputs for LCOS}}

\end{tabular}
\end{adjustbox}
\end{table}

A minimum LCOS metrics as a solely technology design objective is not enough to argue about competitiveness. Regardless of the low or high LCOS indication, the 'variable EP scenario' shows that all included energy storage technologies are valuable. As noted earlier, we define a technology as valuable if it reduces the total system costs. This is the case if a technology is part of an optimised energy system. In Figure \ref{fig:LCOS-Ranges}, all technologies reveal a market potential indicating to be required assets to achieve the minimum total system costs. As a result, instead of improving energy storage by minimising the LCOS, one could maximise the system-value and assess the market potential indicator. Why reducing the total system cost should also be in the interest of technology developers will be discussed in Section \ref{S:MPMusefulness}.


\afterpage{%
\begin{figure*}[htb]
\centering
\includegraphics[trim={0cm 0cm 0cm 0cm},clip,width=0.4\textwidth, angle=-90]{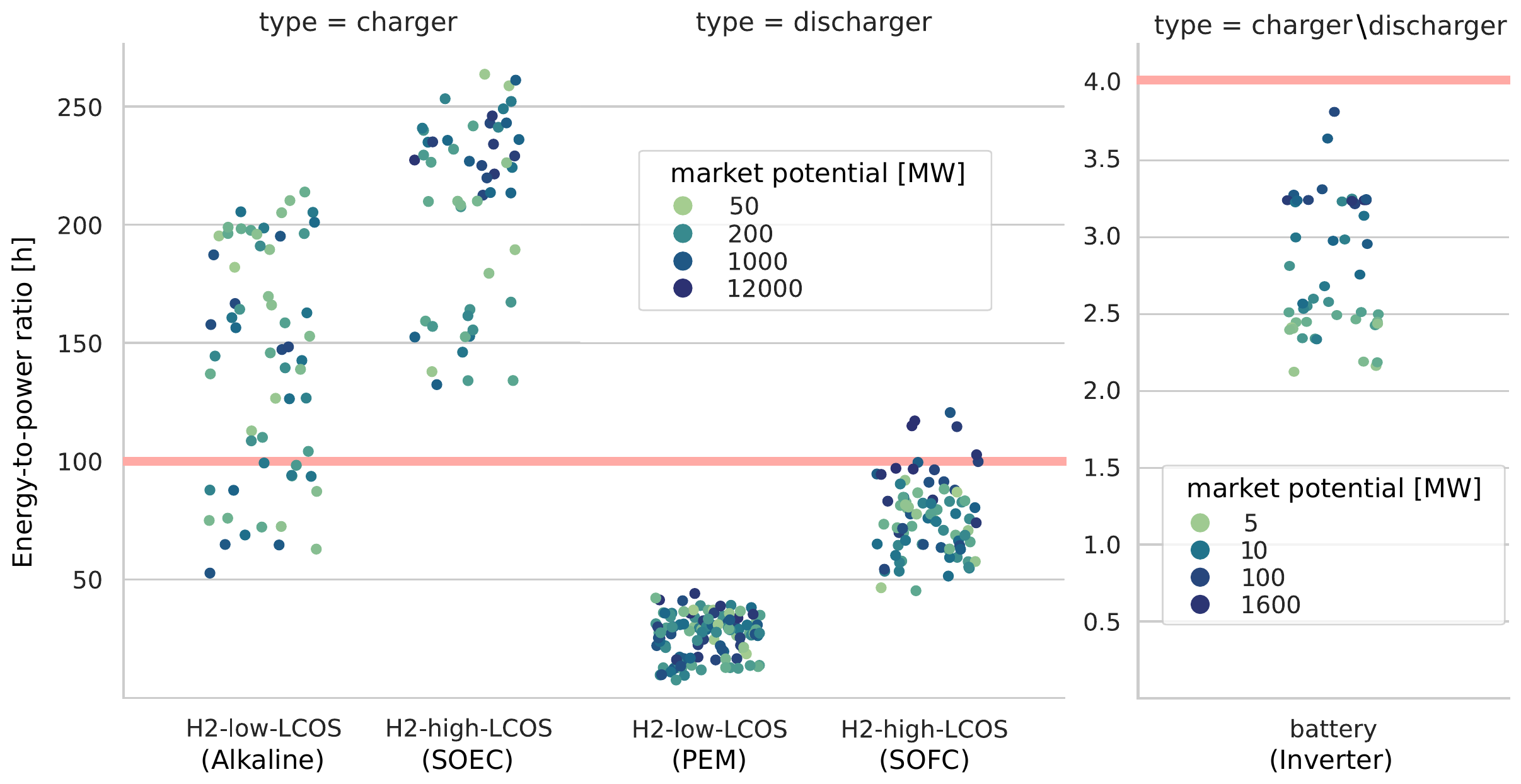}
\caption{Optimal energy to power ratio ranges in the variable EP ratio scenario. The red line represents the fixed EP-ratio scenario assumption. The energy to power ratios are very diversely sized in the 181 buses of the cost-optimal European system layout and in regards to hydrogen and not necessarily equal for charger and discharger. The electrolyser capacity is generally smaller than the fuel cell capacity, which means that slow charging and quick discharge at few moments is desired in the system.}
\label{fig:EP_ratio}
\vspace*{\floatsep}
\centering
\includegraphics[trim=0cm 0cm 0cm 0cm,clip,width=1\textwidth]{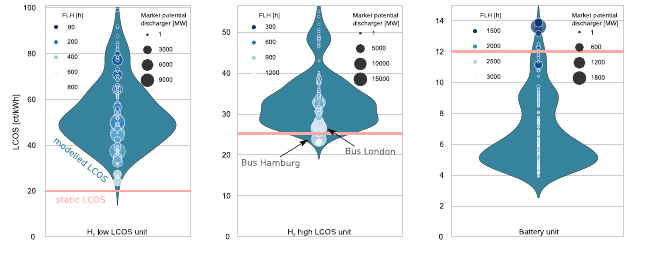}
\caption{Static LCOS results compared to European wide modelled LCOS. The static LCOS is marked by a red horizontal line and was calculated for a set of assumption in Table \ref{table: additional model inputs for LCOS}. In contrast, the modelled LCOS is given as points and uses spatial-temporal dissolved European energy modelling outputs for its calculation. The size of each point shows the optimised market potential of discharger in a given region and helps indicating the relevance. The colour reveals full load hours for each storage technology and helps understanding the operational behaviour which partially lead to the LCOS. The width of the violin plot shows the occurrence in the kernel density estimation, hence, the wider the plot the more buses are located at the respective LCOS cost range. In all cases, buses with less than 1 MW market potential or 80 FLH are removed, keeping the visualisation readable.}
\label{fig:LCOS-Ranges}
\end{figure*}
\clearpage}

\subsection{Market potential method as value indicator} \label{S:MPMapplied}

This section reveals the market potential indicator for each technology and scenario and evaluates it exemplary with the market potential criteria. Exemplary, because as described in Section \ref{S:market potenial method} the MPM scenarios should be chosen according to institutional scenarios or 'beliefs' that might be more likely to impact decision making. As noted earlier, the scenario design of this study is described in Figure \ref{fig:scenario-set-up} and helps to interpret the results.
 
\begin{figure*}[!t]
\centering
\includegraphics[trim={1cm 2cm 6.2cm 2cm},clip,width=0.98\textwidth]{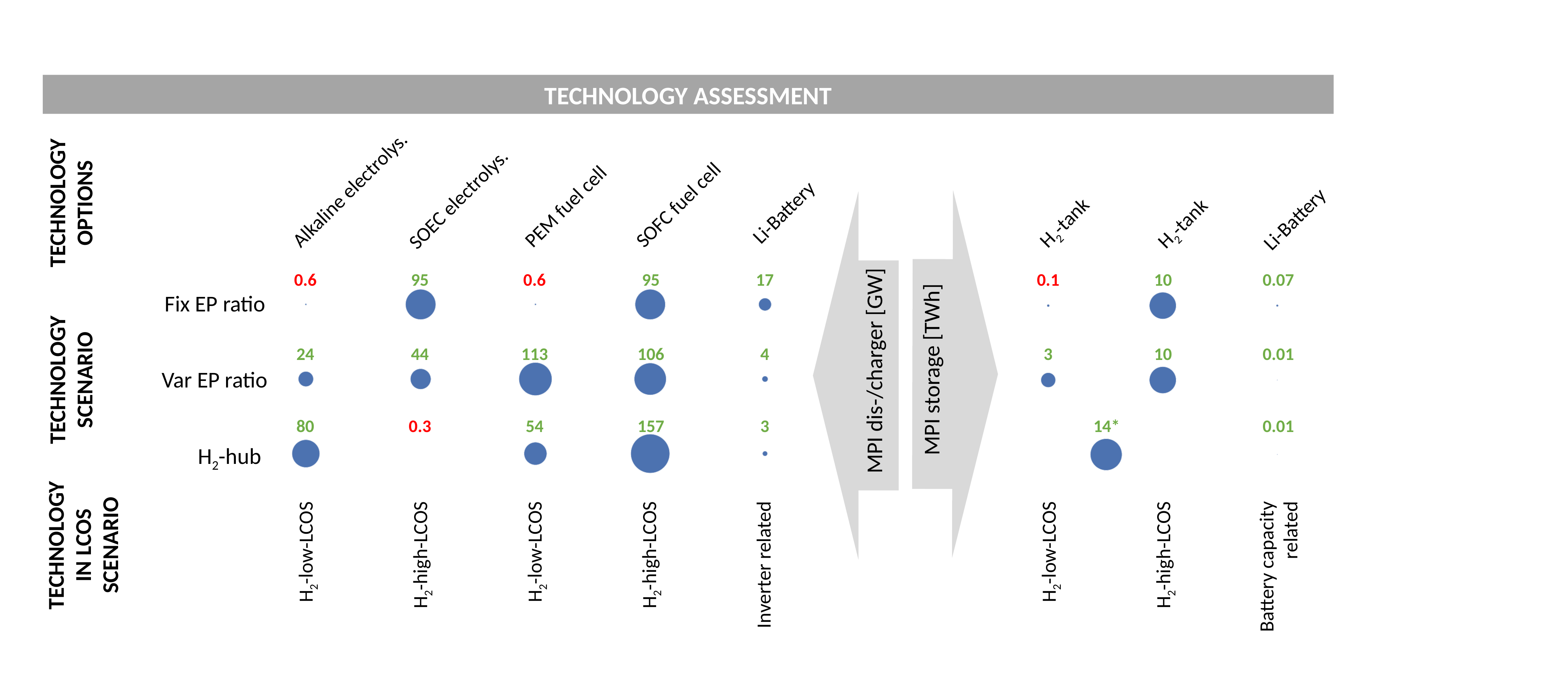}
\caption{Market potential indicator for all charging and discharging components in Europe for three technical storage scenarios in a zero emission electricity system. Despite having the same economic and technical input data the market potential vary drastically between the scenarios. The SOFC fuel cell and Li-battery are according to the market potential method, the technologies which are most likely to be valuable in the exemplary set of scenarios. Because they have an optimised market potential indicator in each scenario. *Refers to the total shared storage capacity. }
\label{fig:market-potential-scatter}
\end{figure*}

Figure \ref{fig:market-potential-scatter} shows the total market potential indicator for all expandable storage components in the European market. How this market potential can be disaggregated over Europe is demonstrated for chargers and the variable EP ratio scenario in Figure \ref{fig:EU-map-electrolyser}.

The first scenario shows a fixed energy to power ratio of 100h (10TWh/95GW) for hydrogen technologies and 4h (0.07TWh/17GW) while the charging and discharging market potential are constrained to be equal for one storage unit. In this scenario, the main optimised hydrogen technology is the high LCOS case of the static LCOS calculation, whereby the low LCOS case reveals a negligible market potential. It means in simple terms that the high LCOS hydrogen unit is more likely to be valuable and worthwhile to design or manufacture due to the approximately two orders of magnitude higher market potential.

In the second scenario, when all hydrogen storage components, and the battery inverter to capacity ratio, are independently scalable, one can observe a noteworthy reduction of the market potential of battery components. This means that flexible scaling of storage technologies can reduce the viable market for batteries. Further, the optimised energy to power ratio impacts the market potential for hydrogen technologies. Now, both high and low LCOS technologies possess a good market potential and seem desirable as complementary technologies. However, the variable sizing of hydrogen components leads to a market potential shift from charger towards discharger components. For a fixed, variable and $H_2-Hub$ scenario, the total amount of hydrogen charger market potential (summing low and high LCOS components) shifts from 95, 68 and 80 GW to a hydrogen discharger market potential of 95, 219 and 211 GW, respectively. This makes the hydrogen discharger components the clear winner of variable sizing through a rough doubling in market potential. 

Concerning the $H_2-Hub$ scenario, when components are variable sized and diverse $H_2$ electrolyser and fuel cell technologies can simultaneously use the same storage tank, then the storage technologies' market potential changes remarkable again. It makes the before well desirable solid oxide electrolyser as technology almost negligible in terms of market potential. 

As a result, the market potential indicator reveals that the design freedom of storage is crucial because it impacts the value assessment. For instance, when variable component sizing is possible, the PEM fuel cell and the Alkaline electrolyser seem to be more desirable while Li-batteries lose importance in the electricity system.  

Applying the full MPM with the market potential criteria leads to the insight that all the implemented storage components can be considered valuable. The value is thereby derived from the fact that at least one scenario possesses a positive market potential indicator. However, only the Li-battery, as well as the SOFC fuel cell, are the most likely valuable technologies as they are optimised in all scenario's and exceed a self-defined 1 GW threshold criteria. As noted earlier, such a threshold might be set by a manufacturer to define a minimal viable market for a technology worth to invest. The knowledge derived from the market potential criteria can lead to implications, for instance, that the Alkaline electrolyser manufacturer can actively mitigate their value risk by promoting variable sizing.

Finally, the presented insights underline the misleading concept of solely cost minimising technologies. Not always a technology with the lowest investment or LCOS is most valuable. It can also be the more expensive technology that can lead to a cheaper future electricity system.

\subsection{The relevance of the market potential method}\label{S:MPMusefulness}

The market potential indicator is a helpful metric from a practical and computer modelling perspective for manufacturers, developers and researchers.
The most important reason for the usefulness is that the market potential is a driver for business. Successful companies want to generate money for their stakeholders and, hence, are driven by two things, growth and profitability. The market potential indicator for a specific product can relate the growth potential to profitability. For instance, when a company expects to offer a future product for net costs of 10 \euro/kWh, it could include these costs in the energy system model with a profit and risk premium of 5 \euro/kWh. The modelling output is the market potential indicator, which is related to the profit and risk premium of 50\%. As a result, the market potential method can be useful for growth and profit evaluations of future storage technology. 

\begin{figure}[ht]
\centering
\includegraphics[trim={0cm 0.5cm 0cm 1cm},clip,width=0.48\textwidth]{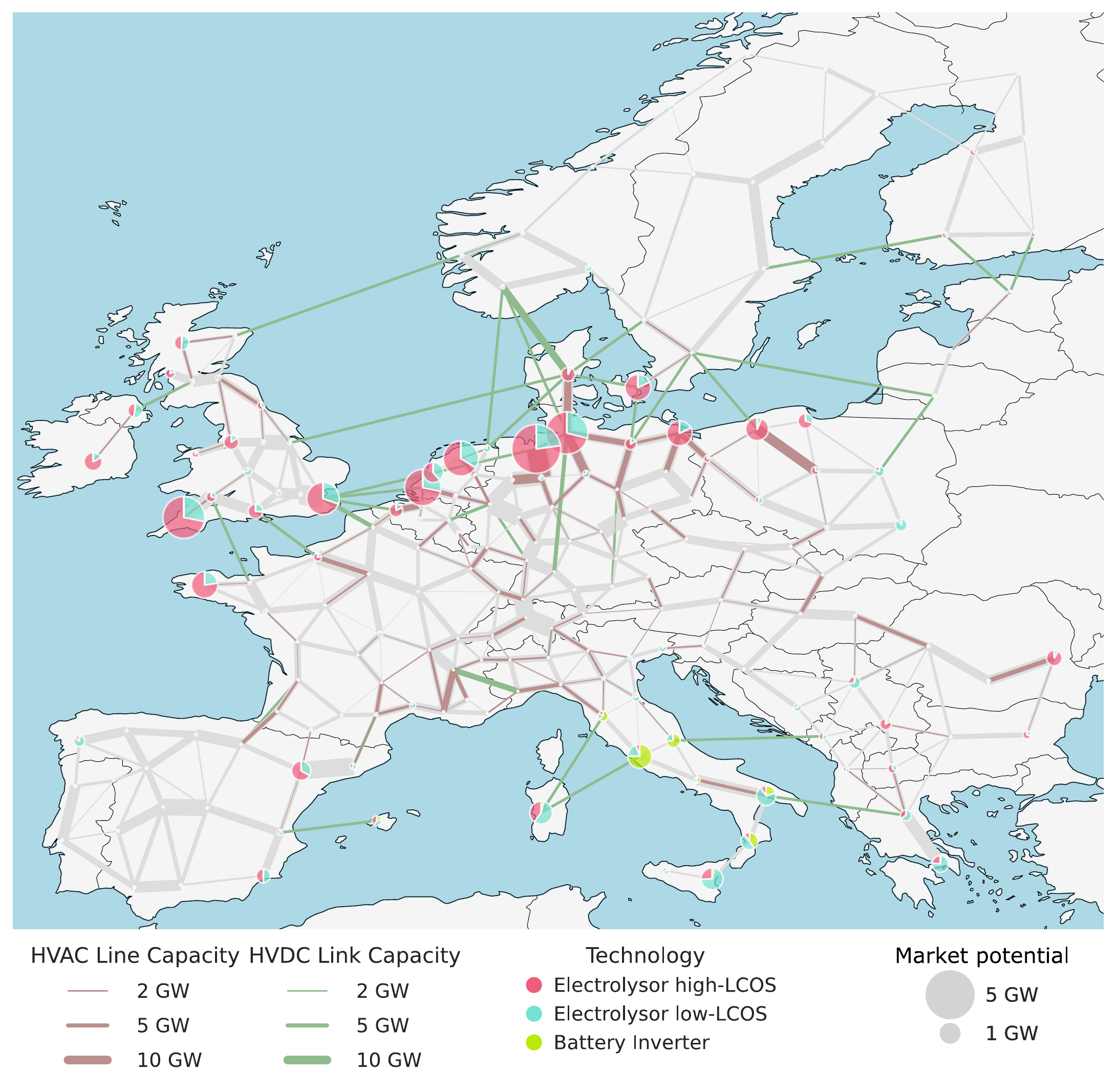}
\caption{Optimal energy storage charger distribution in the variable energy to power sizing scenario. Showing the location of market potential in a 100\% emission reduction scenario. When compared to Figure \ref{fig:EU-map}, most hydrogen units are co-located with wind plants while batteries gravitate towards solar plant optimised areas \cite{Victoria2019TheSystem}.}
\label{fig:EU-map-electrolyser}
\end{figure}

Second, the market potential can give insights into where growth markets are located and for what reason. This can be achieved since the disaggregated market potential can identify regions with future technology expansion (see Figure \ref{fig:EU-map-electrolyser}). The electrolyser distribution reveals that in many locations, high and low LCOS units complement each other. Additionally, when storage components are compared to the generation distribution from Figure \ref{fig:EU-map}, most hydrogen units are co-located at regions with wind plants (mostly northern regions). At the same time, batteries gravitate towards solar plant optimised areas (mostly southern regions). A reason for the observed co-location might be the diurnal solar power pattern and the multi-day to weekly wind power pattern, which creates a network constrained mismatch suitable for the given storage characteristics \cite{Victoria2019TheSystem}. 

Third, the market potential is useful as an indicator of future cost reductions. Because with the market potential, one can assume future technology deployment, which is an implicit factor in learning by doing cost reduction effects \cite{Kittner2017EnergyTransition} or a factor that can be incorporated into process-based cost analysis to evaluate the cost reduction potential \cite{DeSantis2017, James2016}. 

Forth, the market potential can reduce the structural uncertainty of the linear programming energy system model itself. Initial cost assumptions as model inputs are often made without knowing deployment numbers achieved in the optimisation. Nevertheless, it is known that more extensive deployment can reduce costs due to learning effects \cite{Kittner2017EnergyTransition}. Since after the first model run the market potential can function as a cost reduction signal, one can in an iterative or sequential solution approach improve the input accuracy and, hence, lower the structural uncertainty. 

Finally, the operational behaviour can be analysed with the spatially distributed market potential due to the use of energy system models, which gives operational times series of optimised technologies. These time series can be used to identify operational patterns and full load hours, which might be helpful in technology design decisions.

\section{Critical Appraisal}

What the market potential gives its power to resolve the complex value of energy storage - the energy system model - also introduces typical limitations found in this domain. The fundamental challenge of any mathematical energy model is to represent a realistic future energy system that includes all relevant physical, social and political details \cite{Pfenninger2014EnergyChallenges}. Current approaches encounter limitations to represent these details. For instance, models often aggregate in space, time and technological resolution, and ignore unit commitment constraints to reduce the computational requirements at the cost of reduced accuracy to represent future scenarios; or assume perfect and complete markets, where actors have perfect foresight. Both deviate from what can be accomplished in reality \cite{Horsch2018PyPSA-Eur:System}, and as pointed out in the introduction, it can be important to address additional values of energy storage.

These energy model limitations can be understood as (1) structural uncertainty related to the imperfect mathematical description of the physics and (2) parametric uncertainty that refers to imperfect knowledge of input values, i.e. impacted by innovation or behaviour. Both compromise every kind of mathematical model with increasing uncertainty looking into the more distant future and vary from model to model \cite{Bistline2020EnergyNeeds, Neumann2021TheModel, DeCarolis2017FormalizingModelling}. The most important uncertainties of PyPSA-Eur are summarised in \cite{Horsch2018PyPSA-Eur:System}, for instance, that demand profiles for regions in a country are not disaggregated and only scaled by the GDP of the regions, hence, representing not local differences; or missing multi-horizon optimisation, which can help to describe investment pathways and lock-in effects; or the only focus on the electricity system, missing alternative flexibility competitors from other sectors. 

Nevertheless, most of the uncertainties can be reduced by improving future mathematical descriptions of the reality and by strategies to reveal remaining uncertainties \cite{Pfenninger2014EnergyChallenges}. For instance, one compelling way to address parametric uncertainty is to give robust insights about what actions are viable within given cost assumptions by exploring systematically scenarios and the feasibility space near the optimum, such as applied in \cite{Neumann2021BroadNear-Optimality}. An approach to address the structural uncertainty, includes this study's missing energy storage values for sub-hourly grid services and risk confronted investment and operation. In PyPSA-Eur many of these certainty creating features can be implemented in short-term by state of the art techniques.  

In the context of the above-described uncertainties, this study does not seek to reveal the one true future prediction. It instead shows a set of possible future scenarios with different technological design freedoms for the only purpose of comparing different storage design evaluation methods.

Future work can reduce the limitations of this study, such as the inclusion of sector coupling and pathway optimisation. Further, this study considered energy arbitrage under perfect and complete markets. Another branch of work can include more services relevant to grid stability and risk approaches, for instance, by investigating the impact of imperfect and incomplete market conditions and higher spatio-temporal resolutions regarding market potential method results. Finally, what might be valuable in Europe could look different in other regions. Technology developers would benefit from a global value assessment. Therefore, it is of utmost importance to expand open energy system models to cover most parts of earth.

\section{Conclusion}

In the context of storage technology evaluation methods, cost reduction approaches are failing to account for system values. This study observed that most energy storage technologies are designed with the aim to reduce their component or storage system costs ignoring the interaction with the energy system. However, we showed that two hydrogen long-term storages, both cheap and expensive, can simultaneously provide benefits to the wider energy system. Therefore, missing with existing cost reduction approaches values a technology can or cannot provide in a wider energy system might misguide technology innovation.

System-value approaches aim to acknowledge wider energy system benefits, however, existing approaches are not practical in the current design for technology evaluation. In this paper, we overcome many existing limitations with the new introduced market potential method that can be described as a systematic deployment assessment. The market potential method provides a complementary approach to evaluate energy storage technology from a system value perspective.

In summary, the market potential method has implications for practical and modelling relevant insights for manufacturers, developers and research. It can be used to 
\begin{itemize}
\item support technology design-decision making with growth signals of magnitude and location, 
\item improve the technology by changing operational behaviour or adapting material or process selection to be most valuable for the energy system, 
\item concentrate policy endeavours to come closer to perfect market circumstances, or to
\item enhance energy modelling as evaluation tool itself.
\end{itemize}

The new method strongly depends on energy system modelling. Improving energy system model design and reducing uncertainty is essential for a successful adoption. Here it is of unquestionable value to use open data and open source models to build trust and credibility for decisions.

The economist Milton Friedman said that ``there is one and only one social responsibility of business–to use its resources and engage in activities designed to increase its profits so long as it stays within the rules of the [market] game, which is to say, engages in open and free competition without deception or fraud." This might sound convenient in many cases, but in the context of developing energy technology, the 'game' is constantly changing due to the energy transition and sector coupling, aiming at complete and perfect markets. Thus, maybe it is time to look beyond the cost reduction paradigm and short-term profit focus - to develop technology that leads to lower system cost and winning the market of the future. The market potential method could contribute to this.


\nomenclature[A]{\(MPM\)}{Market Potential Method}
\nomenclature[A]{\(MPI\)}{Market Potential Indicator}
\nomenclature[A]{\(SOEC\)}{Solid-Oxide Electrolyser}
\nomenclature[A]{\(SOEF\)}{Solid-Oxide Fuel Cell}
\nomenclature[A]{\(PEM\)}{Proton Exchange Membrane}
\nomenclature[A]{\(VRE\)}{Variable renewable energy sources}
\nomenclature[A]{\(EP\)}{Energy to Power}
\nomenclature[A]{\(LCOS\)}{Levelized Cost of Storage}
\nomenclature[A]{\(GHG\)}{Greenhouse gas}
\nomenclature[A]{\(VRE\)}{Variable Renewable Energy}
\nomenclature[A]{\(HVAC\)}{High Voltage Alternating Current}
\nomenclature[A]{\(HVDC\)}{High Voltage Direct Current}
\nomenclature[A]{\(H2\)}{Hydrogen}
\nomenclature[A]{\(NPV\)}{Net Present Value}
\nomenclature[A]{\(WSB\)}{Whole System Benefit}
\nomenclature[A]{\(IRR\)}{Internal Rate of Return}
\nomenclature[A]{\(ROI\)}{Return of Investment}
\printnomenclature

\section{Declarations}


\section*{Ethics approval and consent to participate - Not applicable}

\section*{Consent for publication - Not applicable}

\section*{Code and Data availability}
Code and data to reproduce results and illustrations are available on {\color{blue}\href{https://github.com/pz-max/beyond-cost}{GitHub}} https://github.com/pz-max/beyond-cost.

\section*{Declaration of Competing Interest}
Author, Aristides Kiprakis is Editorial Board Member of Carbon Neutrality. Beyond this, the authors declare that they have no known competing financial interests or personal relationships that could have appeared to influence the work reported in this paper. 

\section*{CRediT authorship contribution statement} 
Conceptualization: MP; 
Methodology: MP; 
Software: MP; 
Validation: MP; 
Formal analysis: MP; 
Investigation: MP; 
Ressources: MP, AK; 
Data Curation: MP; 
Writing - Original Draft: MP; 
Writing - Review \& Editing: MP, FN, AW, DF, AK; 
Visualization: MP; 
Supervision: MP, DF, AK; 
Project administration: MP, DF, AK; 
Funding acquisition: MP, DF, AK; 
\section*{Acknowledgements} 
This research was supported by UK Engineering and Physical Sciences Research Council (EPSRC) grant EP/P007805/1 for the Centre for Advanced Materials for Renewable Energy Generation (CAMREG). M.P. would like to thank Frank Venter, Emmanuel Paez, Nicole Chen, Mousa Zerai, Martin Kittel, Olukunle Owolabi and Thomas Morstyn for helpful comments and inspiring discussion.

\section{Appendix}\label{Sec: Appendix}

The following paragraphs formulate PyPSA-Eur based on \cite{Horsch2018PyPSA-Eur:System, Neumann2021CostsSystem, Neumann2021TheModel, Brown2018PyPSA:Analysis, Parzen2022AlleviateCycling}.

The objective of PyPSA-Eur is to minimise the total system costs, comprised of annualised capital and operational expenditures. Capital expenditures include capacity-related, long-term investment costs $c$ at location $i$ for generator $G_{i,r}$ of technology $r$, storage energy capacity $H_{i,s}^{store}$, charging capacity $H^+_{i,s}$ and discharging capacity $H^-_{i,s}$ of technology $s$ and transmission line $F_{l}$. Operational expenditures include energy-related variable cost $o$ for generation $g_{i,r,t}$ and storage charging $h^+_{i,r,t}$ and discharging $h^-_{i,r,t}$, as well as energy-level related storage cost $e_{i,s,t}$. Thereby, the operation depends on the time steps $t$ that are weighted by duration $w_{t}$ that sums up to one year  $\sum_{t=1}^{T} w_{t}= 365days*24h = 8760h$.

\begin{equation}
\begin{aligned}
\min_{G,H,F,g,h,e} \quad & \left(\text{Total System Cost}\right) = \\
\min_{G,H,F,g,h,e} \quad &   
\Bigg[ \sum_{i,r}(c_{i,r} \cdot G_{i,r}) 
+
\sum_{l}(c_{l} \cdot F_{l}) \\
\quad & 
+ \sum_{i,s}(c^{store}_{i,s} \cdot H_{i,s}^{store} + c^{-}_{i,s} \cdot H^-_{i,s} + c^+_{i,s} \cdot H^+_{i,s}) \\
\quad & 
+ \sum_{i,r,t}(o_{i,r} \cdot g_{i,r,t} \cdot w_t) + 
\sum_{i,s,t}\big((o^+_{i,s} \cdot h^+_{i,s,t} + o^-_{i,s} \cdot h^-_{i,s,t}) \cdot w_t\big) \\
\quad & 
+ \sum_{i,s,t}(o^{store}_{i,s} \cdot e_{i,s,t} \cdot w_t) \Bigg] \\
\end{aligned}
\end{equation}

The objective function is subject to multiple linear constraints to make scenarios more realistic, leading to a convex linear program with continues variables. The constraints explained in the following in more detail consist of i) demand equals supply constraint, ii) geophysical and operational constraint for generators, storage units as well as power lines, iii) Kirchhoff's current and voltage law constraints that represent the physics of electric energy flows in the power network, iv) a recovering cyclic energy storage constraint and finally, and v) greenhouse gas emissions reduction constraint. 
Such linear problems have in general one unique objective value with sometimes multiple non-unique operational solutions \cite{Parzen2022AlleviateCycling}, making complex problems solvable in reasonable amount of time (sometimes multiple days).

The first constraint requires that for all substations demand equals supply for all times and locations which is needed for stable energy system operation.

\begin{equation}
\begin{aligned}
D_{i,r,t} = S_{i,r,t} \quad \forall i,r,t 
\end{aligned}
\end{equation}

Secondly, since generator and storage units as well as transmission lines can experience geographical restriction, PyPSA-Eur can constrain the installed capacities and gives the options for lower as well as upper limits. 

\begin{equation}
\begin{aligned}
\underline{G}_{i,r} \leq G_{i,r} \leq \overline{G}_{i,r} \quad \forall i,r 
\end{aligned}
\end{equation}

\begin{equation}
\begin{aligned}
\underline{H}_{i,s} \leq H_{i,s} \leq \overline{H}_{i,s} \quad \forall i,s 
\end{aligned}
\end{equation}

\begin{equation}
\begin{aligned}
\underline{F}_{l} \leq F_{l} \leq \overline{F}_{l} \quad \forall l 
\end{aligned}
\end{equation}

Such constraints help to implement social, environmental or physical based boundary conditions. Atlite is one of the tools that are implemented in PyPSA to quantify for instance the land availability for solar and wind power plants by incorporating protected areas and land coverage classification data to reduce the renewable installation potential \cite{Hofmann2021Atlite:Series}.

Thirdly, while the previous constraint only limits the installations, some energy system components require time-varying operational limits. Examples for such technologies are renewable generators and power lines with dynamic line-rating (DLR) which operation highly depend on the weather signals. With roughly 20x20km globally rasterized era5 weather data that are available for the last 30 years, again produced by Atlite, PyPSA-Eur can limit the rated power of generators $G_{i,r}$ and lines $F_{l}$ by a location and time dependent variable, i.e. temperature, wind speed, humidity and solar irradiation, such that

\begin{equation}
\begin{aligned}
0 \leq 
g_{i,r,t} \leq 
\overline{g}_{i,r,t} G_{i,r}
\quad \forall i,r 
\end{aligned}
\end{equation}

\begin{equation}
\begin{aligned}
0 \leq 
f_{l,t} \leq 
\overline{f}_{l,t} F_{l}
\quad \forall i,r 
\end{aligned}
\end{equation}

Thirdly, the PyPSA-Eur model typically includes a linearised power flow constraint modelling the physicality of the power transmission network. A very distinctive feature compared to most other energy system planning models \cite{Brown2018PyPSA:Analysis}.  This is done by including Kirchhoff’s Current Law and Kirchhoff’s Voltage Law constraints.

Kirchhoff’s Current Law requires local generators and storage units as well as incoming or outgoing flows $f_{l,t}$, of incident transmission lines described by $K_{i,l}$ as the networks' incidence matrix, to balance the inelastic electricity demand $d_{i,t}$ at each location $i$ and time step $t$

\begin{equation}
\begin{aligned}
\sum_{r}{g_{i,r,t}} + \sum_{s}{h^{-/+}_{i,s,t}} + \sum_{l}{K_{i,l} \cdot f_{l,t}} = d_{i,t} \quad \forall i,s,r,t 
\end{aligned}
\end{equation}

While Kirchhoff’s Current Law accounts for both, AC and controllable DC lines, the Kirchhoff’s Voltage Law only additionally constraints AC power lines. Here the voltage angle difference around every closed cycle in the network must add up to zero. PyPSA-Eur formulates this constraint using linearised load flow assumptions, in particular, cycle basis $C_{l,c}$ of the network graph where the independent cycles $c$ are expressed as directed linear combinations of lines \cite{Horsch2018LinearFlows}. This leads to the constraints  

\begin{equation}
\begin{aligned}
\sum_{l}{C_{l,c}} \cdot
x_l \cdot
f_{l,t}
= 0 \quad
\forall l,t 
\end{aligned}
\end{equation}

where $x_l$ is the series inductive reactance of line $l$ \cite{Neumann2021CostsSystem}. As might be noted, the linearised powerflow assumptions completely disregard the resistance. These assumptions introduce negligible errors when (i) the reactance is much larger than the resistance, such as for high voltage lines, and (ii) the voltage angel differences are small i.e. $sin(\delta) = \delta$ \cite{Horsch2018LinearFlows}.

Fourth, describing storage constraints. Storage charging $h_{i,s,t}^+$ and discharging $h_{i,s,t}^-$ are both positive variables and limited by the installed capacity $H_{i,s,t}^+$ and $H_{i,s,t}^-$.

\begin{equation}
\begin{aligned}
0 \leq h_{i,s,t}^+ \leq H_{i,s}^+ \quad \forall i,s,t 
\end{aligned}
\end{equation}

\begin{equation}
\begin{aligned}
0 \leq h_{i,s,t}^- \leq H_{i,s}^- \quad \forall i,s,t
\end{aligned}
\end{equation}

This formulation keeps the feasible solution space convex, though does not prevent simultaneous charging and discharging, which is often an unrealistic effect that can heavily distort modelling results in net-zero scenarios. Setting adequate variable cost parameter solves this modelling artefact while keeping the problem formulation linear \cite{Parzen2022AlleviateCycling}.

The storage energy level $e_{i,s,t}$ is the result of a balance between energy inflow, outflow and self-consumption. Additional to directed charging and discharging with its respective efficiencies $\eta_{i,s,+}$ and $\eta_{i,s,-}$, natural inflow $h_{i,s,t}^{inflow}$, spillage $h_{i,s,t}^{spillage}$ as well as standing storage losses that reduces the storage energy content of the previous time step by a factor of $\eta_{i,s,+}$ are considered.

\begin{equation}
\begin{aligned}
e_{i,s,t} \quad & = \eta_{i,s,+} \cdot e_{i,s,t-1} + \eta_{i,s,+} \cdot w_t \cdot h_{i,s,t}^+ - \eta_{i,s,-}^{-1} \cdot w_t \cdot h_{i,s,t}^- \\
\quad & + w_t \cdot h_{i,s,t}^{inflow} - w_t \cdot h_{i,s,t}^{spillage} \quad \forall i,s,t
\end{aligned}
\end{equation}

The amount of energy that can be stored is limited by the energy capacity of the installed store unit $H_{i,s}^{store}$ [MWh], which allows independent storage component scaling. 

\begin{equation}
\begin{aligned}
0 \leq e_{i,s,t} \leq H_{i,s}^{store} \quad \forall i,s,t 
\end{aligned}
\end{equation}

To fix the storage technology design, a technology-specific energy to discharging power ratio $\overline{T}_s$ can be multiplied with the capacity of the discharging unit $H_{i,s}^-$ 

\begin{equation}
\begin{aligned}
0 \leq e_{i,s,t} \leq \overline{T}_s \cdot H_{i,s}^- \quad \forall i,s,t
\end{aligned}
\end{equation}

\noindent to define the upper energy limit per installed storage.

Further, the energy storage units are assumed to be cyclic, i.e., the state of charge at the first and last period of the optimization period $T$ (i.e. 1 year) must be equal:

\begin{equation}
\begin{aligned}
e_{i,s,0} = e_{i,s,T} \quad \forall i,s
\end{aligned}
\end{equation}

This cyclic definition is not mandatory but helps with the comparability of model results. It further avoids the free use of storage energy endowment, meaning that the model could prefer to start with a higher and end with a lower storage level to save costs. 

Finally, PyPSA-Eur can constrain the total emissions. These emissions are tracked by a variable at each generator unit, which depends on the supply source or carrier $q$. Allowing to constrain the total emission by a limiting parameter $\overline{GHG}$ by

\begin{equation}
\begin{aligned}
g_{i,r,t,q} \leq \overline{GHG} \quad \forall i,r,t,q
\end{aligned}
\end{equation}

\bibliographystyle{elsarticle-num-names}
\bibliography{references.bib}

\end{document}